\documentclass[aps,prl,twocolumn,amsfonts,longbibliography,citeautoscript,superscriptaddress]{revtex4-1}
\usepackage{xcolor}
\usepackage{bm}
\usepackage{bbold}
\usepackage{amsmath}
\usepackage{amssymb}
\usepackage{epsfig}
\usepackage{verbatim}
\usepackage[T1]{fontenc}
\usepackage{array}

\newcommand{\la}{\langle}
\newcommand{\lla}{\la\!\la}
\newcommand{\ra}{\rangle}
\newcommand{\rra}{\rangle\!\rangle}

\newcommand{\p}{\partial}
\newcommand{\eel}{l}
\newcommand{\rd}{{\rm d}}

\usepackage[colorlinks=true,bookmarks=false,citecolor=blue,linkcolor=blue,hyperfootnotes=true,urlcolor=blue]{hyperref}

\begin{document}

\title{Quantum quench in a driven Ising chain} 
\author{Neil J. Robinson}
\email{neil.joe.robinson@gmail.com}
\affiliation{Institute for Theoretical Physics, University of Amsterdam, Postbus 94485, 1090 GL Amsterdam, The Netherlands}
\altaffiliation{Present address: UKRI EPSRC, Polaris House, North Star Avenue, Swindon SN2 1ET, United Kingdom}

\author{Isaac P\'erez Castillo}
\email{feymanncool@gmail.com}
\affiliation{Departamento de F\'isica, Universidad Autónoma Metropolitana-Iztapalapa, San Rafael Atlixco 186, Ciudad de México 09340, Mexico}

\author{Edgar Guzm\'an-Gonz\'alez}
\email{egomoshi@gmail.com}
\affiliation{London Mathematical Laboratory, 8 Margravine Gardens, London W6 8RH, United Kingdom}

\date{\today}

\begin{abstract}
We consider the Ising chain driven by oscillatory transverse magnetic fields. For certain parameter regimes, we reveal a hidden integrable structure in the problem, which allows access to the \textit{exact time-evolution} in this driven quantum system. We compute time-evolved one- and two-point functions following a quench that activates the driving. It is shown that this model does not heat up to infinite temperature, despite the absence of energy conservation, and we further discuss the generalization to a family of driven Hamiltonians that do not suffer heating to infinite temperature, despite the absence of integrability and disorder.  The particular model studied in detail also presents a route for realising exotic physics (such as the E8 perturbed conformal field theory) via driving in quantum chains that could otherwise never realise such behaviour.  In particular we  numerically confirm that the ratio of the meson excitations masses is given by the golden ratio.
\end{abstract}

\maketitle

{\it Introduction.}---Over the last decade, the nonequilibrium dynamics of quantum systems has attracted a great deal of attention~\cite{gogolin2016equilibration,dAlessio2016from,calabrese2016quantum,cazalilla2016quantum,caux2016quench,essler2016quench,bernard2016conformal,prosen2016quasilocal,vasseur2016nonequilibrium,deluca2016equilibration,langen2016prethermalization,vidmar2016generalized}, motivated by the desire to address fundamental questions: When and how do quantum systems relax to equilibrium? How does one describe this equilibrium? What influences the dynamics and equilibration? Understanding these issues is important when developing descriptions of a growing number of experiments that examine nonequilibrium dynamics, both in cold atomic gases~\cite{polkovnikov2011nonequilibrium,langen2015ultracold} and the solid state~\cite{bovensiepen2012elementary,beye2013time}. The insights gained may play an important role in the development of quantum computing resources, especially when considering how to protect quantum information from the scrambling associated with thermalization.

Recently attention has turned to understanding driven quantum systems, partially due to the realization that such systems can host interesting topological phases (see, e.g., Refs.~\cite{vonkeyserlingk2016phase,vonkeyserling2016phaseII,else2016classification,potter2016classification,roy2016abelian,roy2017periodic,roy2017floquet}) and other exotic behaviors (such as time crystal phases~\cite{khemani2016phase,vonkeyserling2016absolute,zhang2017observation,choi2017observation,khemani2017defining,yao2017discrete,khemani2019brief}). These studies have generated much discussion of how to extend and apply the concepts of equilibrium statistical mechanics in the presence of driving. A particular issue is that, generically, driven quantum systems do not conserve energy. As a result, in the long time limit entropy maximization leads them to heat up to infinite temperature, leading to trivial ergodic behavior. As a result, quantum information is completely scrambled~\cite{dalessio2014longtime,lazarides2014equilibrium,ponte2014periodically}.  Routes to avoid this behavior include introducing disorder to induce a many body localization transition (see, e.g., Refs.~\cite{ponte2015manybody,lazarides2015fate,abanin2016theory,khemani2016phase}), or to consider models that are, in some sense, integrable~\cite{gritsev2017integrable}. 

In this Letter, we consider a driven model that at each point in time is nonintegrable but nonetheless possesses  the dynamics which is governed by a hidden integrability. Using this, we compute the nonequilibrium dynamics of equal-time correlation functions following a quench in which the driving is initiated. The method for attacking this problem can be generalized to a (infinite) family of Hamiltonians, opening the door for future nonperturbative, exact studies. We will see that this whole family of driven quantum systems,  each of which is generically nonintegrable, does not undergo heating to infinite temperature. We will also see that breaking the special structure of this family leads to thermalization to infinite temperature.  

{\it The driven Ising chain.}---We consider a one-dimensional spin-1/2 Ising magnet, driven by oscillatory transverse fields. The Hamiltonian reads  
\begin{equation}
  \begin{split}
    H(t) =& -J \sum_{l=1}^L \sigma^z_l \sigma^z_{l+1} + h^z \sum_{l=1}^{L} \sigma^z_l \\
    &-g\sum_{l=1}^{L}  \Big(e^{-i\Omega t}\sigma ^+_l + e^{i\Omega t}\sigma^-_l \Big),\label{Ht}
  \end{split}
\end{equation}
with $J > 0$ the Ising exchange parameter, $h^z$ a static longitudinal field, $g$ the strength of the transverse fields, which oscillate at frequency $\Omega$, and $L$ the system size. The spin operators $ \sigma^\alpha_l$ act at the $l$th site of the lattice, $\sigma^\pm_l=(\sigma^x_l\pm i \sigma^y_l)/2$,  and we impose periodic boundary conditions $\sigma^\alpha_{L+1} = \sigma^\alpha_1$.  The Hamiltonian~\eqref{Ht} is periodic in time $H(t)=H(t+T)$ with period $T = 2\pi/\Omega$, and could be realized in the quasi-1D ferromagnet CoNb$_2$O$_6$~\cite{coldea2010quantum,robinson2014quasiparticle} by application of oscillating transverse fields. 

At a generic time, the Hamiltonian consists of an Ising interaction term and fields in all ($x,y,z$) directions. Thus instantaneously the Hamiltonian $H(t)$ is \textit{nonintegrable}, and the exact computation of quantities seems unlikely. In the following we will see that this is in fact not the case -- there exists a hidden integrable line within this model where exact results can be obtained. Furthermore, away from this  integrability we will draw general insights.

{\it Time evolution of observables.}---We will now consider how a state $|\Psi_0\ra$ evolves under the Hamiltonian~\eqref{Ht} at times $t>0$. The time-evolved state $|\Psi(t)\ra$  will be a solution of the time-dependent Schr\"odinger equation
\begin{equation}
  \Big[ i\hbar \p_t - H(t) \Big] |\Psi(t)\ra = 0, \label{SchrodingerEq}
\end{equation}
subject to the initial condition $|\Psi(t=0)\ra = |\Psi_0\ra$. Herein we set $\hbar=1$, which defines our units. The formal solution of Eq.~\eqref{SchrodingerEq} is well-known:
\begin{equation}
|\Psi(t)\rangle = \mathbb{T} \exp\left( -i \int_0^t \mathrm{d}t'\, H(t') \right) |\Psi_0\rangle, 
\end{equation}
however, using this to compute time-evolution is a challenge due to the explicit time-ordering ($\mathbb{T}$) of the exponential. To make some headway on this problem we apply a time-dependent unitary transformation $U(t)$~\footnote{As discussed in Refs.~\cite{kolodrubetz2013classifying,kolodrubetz2017geometry,weinberg17adiabatic}, there is nice geometric interpretation of unitary transformations that depend on a continuous parameter (e.g., time $t$) in terms of gauge potentials.}, multiplying both sides of Eq.~\eqref{SchrodingerEq} from the left by $U(t)$ and inserting a factor of $\mathbb{1} = U(t)^\dagger U(t)$ between the wave function and the operators:
\begin{equation}
U(t) \Big[ i \p_t - H(t) \Big] U^\dagger(t) U(t) |\Psi(t)\ra  = 0.
\end{equation}
The problem can become much simpler if there is a choice of $U(t)$ such that this reduces to an effective \textit{time-independent} Schr\"odinger equation. Choosing~\cite{takayoshi2014magnetization} 
\begin{equation}
U(t) = \exp\Big(\frac{i\Omega t}{2} \sum_l \sigma^z_l\Big) \equiv e^{\frac{i\Omega t}{2} \sigma^z_{\rm tot} }, \label{Ut}
\end{equation}
we map Eq.~\eqref{SchrodingerEq} to a time-independent Schr\"odinger equation, $( i \p_t - H_{\rm st})|\Phi(t)\ra = 0$, with an effective static Hamiltonian
\begin{equation}
H_{\rm st} = \sum_{l=1}^L \left[ -J \sigma^z_l \sigma^z_{l+1} + \left(h^z-\frac{\Omega}{2}\right) \sigma^z_l - g \sigma^x_l \right]. \label{Hst1}
\end{equation}
The wave function transforms as $|\Phi(t)\ra = U(t) |\Psi(t)\ra$. This reduction to a static problem is not evident in the Magnus expansion \footnote{See the Supplemental Material, which also contains the references \cite{blanes2009magnus,jordan1928uber,fisher1968toeplitz,auyang1974theory,basor1991fisherhartwig,forrester2004applications,bottcher2006szego,karlovich2007asymptotics,deift2011asymptotics}, for: (i) A discussion of the Magnus Expansion for the driven problem considered; (ii) details of how two-point correlation functions transform under action of the time-dependent unitary transformation; (iii) a discussion of the full-time evolution of observables for a special quench; (iv) detailed derivations of the required ``sudden quench'' correlation functions; (v) details of the numerical algorithm for computing time-evolution of the driven system and a numerical check of the absence of heating to infinite temperature outside the integrable line.}.

Diagonalizing~\eqref{Hst1} to obtain eigenstates $|E_n\ra$  with energies $E_n$, the time-evolved state can be written as
\begin{equation}
|\Psi(t)\ra = \sum_n \exp\Big[-i\left(E_n + \frac{\Omega}{2} \sigma^z_\text{tot}\right)t\Big] |E_n\ra \la E_n |\Psi_0\ra\, .\label{Psit}
\end{equation}
The states $|E_n\ra$ are \textit{not eigenstates} of $\sigma^z_\text{tot}$ and thus each term in Eq.~\eqref{Psit} undergoes nontrivial dynamics. While~\eqref{Psit} is highly nontrivial, there is no need to despair. Our problem reduces to a tractable one if we focus on equal-time correlation functions, as one can use that the operator $U(t)$ acts in a simple manner on the spin operators: 
\begin{equation}
  \begin{split}
    U(t) \sigma^{x,y}_l U(t)^\dagger &= \cos(\Omega t) \sigma^{x,y}_l \mp \sin(\Omega t) \sigma^{y,x}_l,\\ 
    U(t) \sigma^z_l U(t)^\dagger &= \sigma^z_l.
  \end{split}
  \label{transformation}
\end{equation}

\textit{Mapping to a ``sudden quench''.---}Let us now consider the time-evolution of one-point functions $s^\alpha(t) = \la \Psi(t) | \sigma^\alpha_l |\Psi(t)\ra$, where the result
is independent of $l$ by translational invariance.
Using~\eqref{transformation} these become
\begin{equation}
  \begin{split}
    s^z(t) &= s^z_{\text{st}}(t), \\
    s^x(t) &= \cos(\Omega t)  s^x_{\text{st}}(t) - \sin(\Omega t) s^y_{\text{st}}(t), \\
    s^y(t) &= \cos(\Omega t) s^y_{\text{st}}(t) + \sin(\Omega t) s^x_{\text{st}}(t). \\
  \end{split}
  \label{onepts}
\end{equation}
Here each time-dependent expectation value on the right hand side describes time-evolution induced by a sudden quench to the static Hamiltonian~\eqref{Hst1} when starting from the initial state $|\Psi_0\rangle$:
\begin{equation}
s^\alpha_{\text{st}}(t) = \sum_{n,m} e^{i(E_n-E_m)t}  \la \Psi_0| E_n\ra\la E_n| \sigma^\alpha_l |E_m\ra\la E_m | \Psi_0\ra.  
\end{equation}
Equations similar to~\eqref{onepts} can be written for the two-point functions, $s^{\alpha\beta}(\ell;t) = \la \Psi(t) | \sigma^\alpha_j \sigma^\beta_{j+\ell} |\Psi(t)\ra$. 
These are tractable, but a little unwieldy, so are given in~\cite{Note2}. All time-evolved correlation functions are reduced to oscillatory factors multiplying ``sudden quench'' correlation functions. Thus for this driven problem, we can apply the techniques developed for sudden quantum quenches to compute the time-evolution of observables. 

Having reduced the problem from one with driving to an effective sudden quench, let us return to the static Hamiltonian~\eqref{Hst1}. This describes a quantum Ising chain with both transverse $g$ and longitudinal $h = h^z-\Omega/2$ fields. The two fields can be independently controlled via the amplitude $g$ and frequency $\Omega$ of the driving, see Eq.~\eqref{Ht}. Two interesting cases are immediately apparent. Firstly, if the frequency of the driving is tuned to a  $\Omega = 2 h^z$, the longitudinal field is removed from the static Hamiltonian, which then describes the integrable quantum Ising chain~\cite{pfeuty1970onedimensional}. Secondly, one can consider tuning both the amplitude and the frequency such that  $g=J$ and $|h^z-\Omega/2| \ll g$,
where one realizes the lattice limit of the exotic $E_8$ perturbed Ising conformal field theory~\cite{zamolodchikov1989integrals} (which has recently received renewed attention thanks to its nonthermal properties~\cite{rakovsky2016hamiltonian,hodsagi2018quench,james2019nonthermal,robinson2019signatures}, despite an absence of integrability). In this work, we will focus on the first scenario and describe the full time-evolution of one- and two-point functions in this driven problem. We will touch upon the second case towards the end.

When $\Omega=2h^z$, the static Hamiltonian reads:
\begin{equation}
  H^0_{\text{st}}= -J\sum_{l=1}^L \sigma^z_l \sigma^z_{l+1} - g \sum_{l=1}^L \sigma^x_l\,. 
  \label{H0st}
\end{equation}
This is the quantum Ising chain, which can be mapped to free fermions and so is exactly solvable~\cite{pfeuty1970onedimensional}. This reveals that, along the line  $\Omega=2h^z$, there is a hidden integrability in the problem (despite, instantaneously, the Hamiltonian $H(t)$ being nonintegrable). Sudden quenches in the transverse field Ising model have been extensively studied, with many exact results being known, see in particular the works of Calabrese, Essler and Fagotti~\cite{calabrese2011quantum,calabrese2012quantum,calabrese2012quantumII}. We will exploit some of these results, alongside some new ones, to \textit{analytically} compute the dynamics of observables starting from an initial state $|\Psi_0\ra$ that is then time-evolved with the driven Hamiltonian~\eqref{Ht}. The derivation of these results is rather technical, so we provide the details in the Supplemental Material~\cite{Note2}. 

\textit{Time-evolution in the driven model.---}
Let us now present the time-evolution of correlation functions in the driven model~\eqref{Ht} governed by the effective static Hamiltonian~\eqref{H0st}. We compare our analytical results to numerical results obtained on small finite lattices (our numerical algorithm is explained in~\cite{Note2}). 

\begin{figure}
  \begin{tabular}{l}
  (a) \\
  \includegraphics[width=0.5\textwidth]{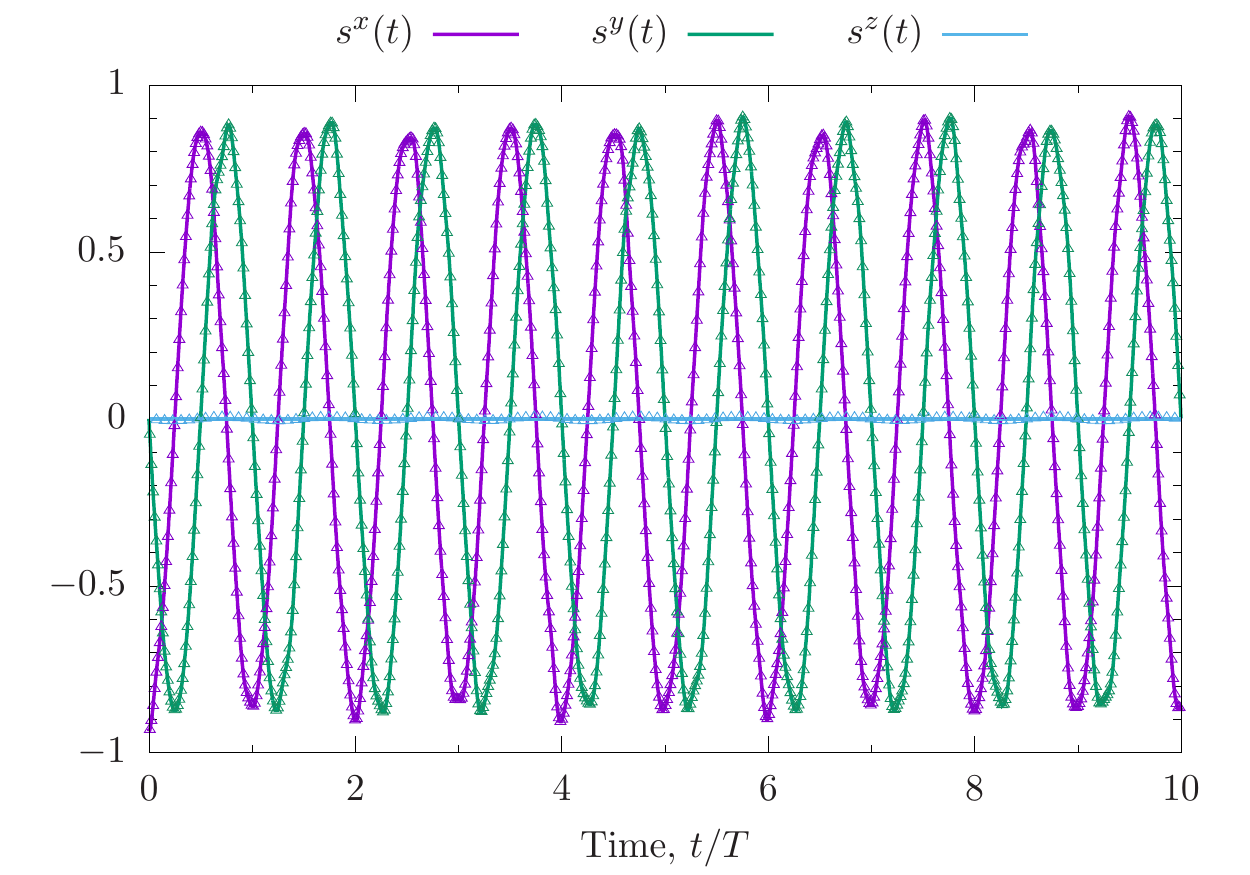}\\
  (b) \\
  \includegraphics[width=0.5\textwidth]{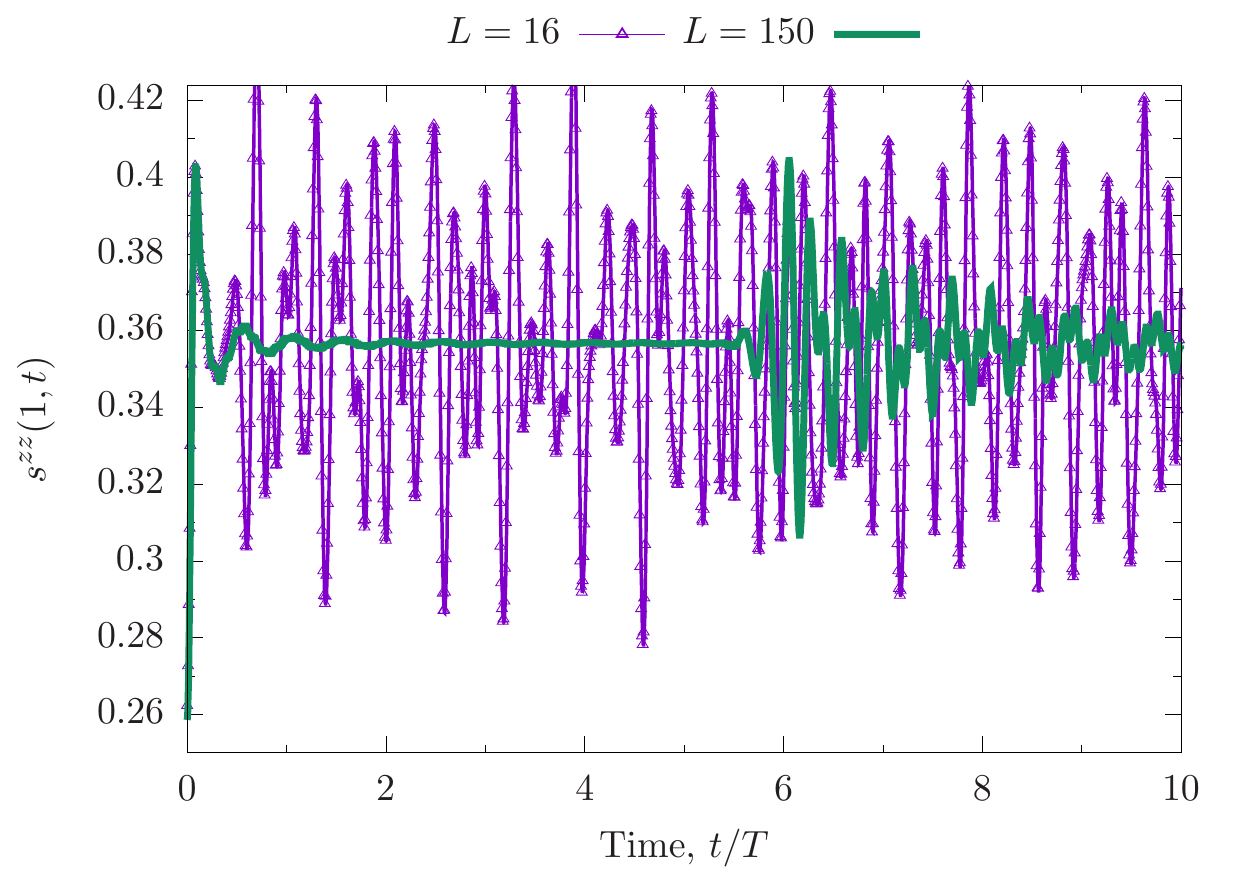}\\
  \end{tabular}
  \caption{(a) Time-evolution of one-point functions starting from the ground state of the quantum Ising chain, $H(t=0)$ with $J = 1, h^z = 0 , g = 2$ and time-evolved with the driven Hamiltonian $H(t)$~\eqref{Ht} with $J = 1, \Omega = 1, h^z =0.5, g = 1.5$ for a system with $L=16$ sites. The behaviour remains the same for bigger system sizes and larger times. Lines represent analytical results, while points show numerically exact time-evolution. (b) Time evolution of the two-point function  $s^{zz}(1,t)$ of two adjacent sites for the same quench, for different system sizes  $L$. We see that $s^{zz}(1,t)$ converges to a non-zero value. The revival of the fluctuations is a finite size effect, as can be seen by increasing the system size.}
 \label{fig:timeevo}
\end{figure}

In Fig.~\ref{fig:timeevo} we present the results for one- and two-point functions for a particular quench. We see that the one-point functions synchronize to the driving frequency $\Omega$ and no heating to infinite temperature occurs, not even if we restrict the study of the system to stroboscopic times. For the two-point functions, we see that $s^{zz}(1,t)$ converges to a non-zero stationary value, confirming the absence of infinite heating.  Although not included in the figure, we mention that the remaining two-point functions synchronize to the period $\Omega$  like the one-point functions, except for $s^{yz}(1,t)$ and $s^{xz}(1,t)$ that converge to zero.

A particularly simple, and solvable in closed form, scenario is realized when $H_{\rm st}$ coincides with the initial Hamiltonian $H(t<0)$. In this case the ``sudden quench'' correlation functions in expressions such as~\eqref{onepts} reduce to \textit{equilibrium correlation functions}, known since the seminal works of~\textcite{barouch1970statisticalI} and~\textcite{barouch1971statisticalII,barouch1971statisticalIII} in the 1970s. Detailed results in this case are presented in~\cite{Note2} and are, to our knowledge, some of the few closed form exact results known for correlation functions in models with driving. 

\textit{Absence of heating to infinite temperature.---}With observables mapping in a simple manner to those from a sudden quench, it is immediately clear that the system cannot undergo heating to infinite temperature, as is usually assumed to occur in driven systems~\cite{dalessio2014longtime,lazarides2014equilibrium,ponte2014periodically}). This is easily seen for observables that feature only $ \sigma^z_l$ operators, which map exactly to ``sudden quench'' observables (see, e.g., the first line of Eq.~\eqref{onepts}). The long-time limit of observables after a sudden quench will be described via the relevant statistical ensemble; for the case detailed above this is the generalized Gibbs ensemble~\cite{rigol2007relaxation,ilievski2015complete,vidmar2016generalized}. Generically, when $H_{\text{st}}$ is nonintegrable, this will be a finite-temperature Gibbs ensemble~\cite{rigol2008thermalization}. It is worth noting that the absence of heating to infinite temperature is not as result of integrability, but instead is due to the structure of the driving term.  In  Fig.~S1 of~\cite{Note2}, we show an explicit example of a non-integrable system with absence of heating to infinite temperature by working outside the integrable line $\Omega=2h^z$.

We can then ask, what happens if this structure is broken such that we do not map to an effective sudden quench problem? We then expect that in the long time limit the system thermalizes to infinite temperature, due to the absence of both energy conservation and the mapping to a sudden quench problem, combined with entropy maximization. We can examine this numerically by adding terms to our Hamiltonian~\eqref{Ht}, for example:
\begin{equation}
  H_X(t) = H(t) + J_X\sum_{l=1}^L \sigma^x_{l}\sigma^x_{l+1}. \label{Hx}
\end{equation}
The added term breaks $\sigma^z$ conservation, and thus evolves non-trivially under the transformation $U(t)$. This breaks the mapping to a static Hamiltonian, and hence we expect heating to infinite temperature. It is worth noting that the thermalization time scale in Floquet systems can be very large, see e.g. Refs.~\cite{abanin2015exponentially,abanin2017rigorous,else2017prethermal,machado2017exponentially}. (The Floquet model studied in Ref.~\cite{machado2017exponentially} bears some similarity to~\eqref{Hx}.)

\begin{figure}
 \includegraphics[width=0.5\textwidth]{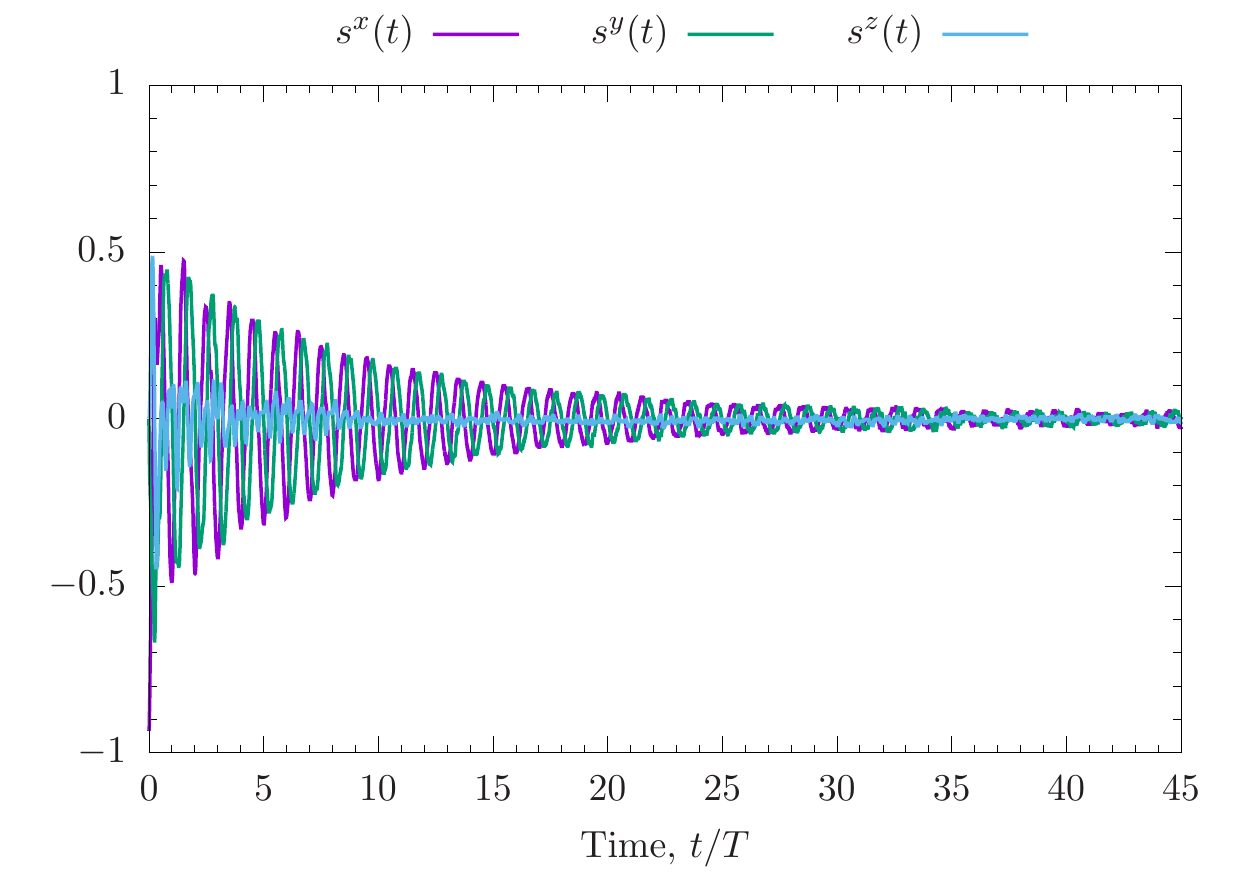}
 \caption{Numerically exact time-evolution  of one-point functions with the driven Hamiltonian $H_X(t)$, Eq.~\eqref{Hx}. This shows that, at the level of one-point functions, breaking the structure of the drive leads to thermalization to infinite temperature, as all the expectation values converge to zero.  The parameters considered were the ones of Fig.~\eqref{fig:timeevo} with $J_x=0.5$, and a Chebyshev expansion of order 64 with a time step  $\Delta t=$0.001 (see~\cite{Note2} for the details of the numerical algorithm employed).}
 \label{fig:inftemp}
\end{figure}

 In Fig.~\eqref{fig:inftemp} we present the time-evolution of one-point functions in the driven model~\eqref{Hx}. With the addition of the $J_X$ term, we see that the system evolves towards a state with $\lim_{t\to\infty}\langle \sigma^\alpha_j(t) \rangle = 0$, corresponding to infinite temperature, at least at the level of one-point subsystems.

\begin{figure}
\includegraphics[width=0.5\textwidth]{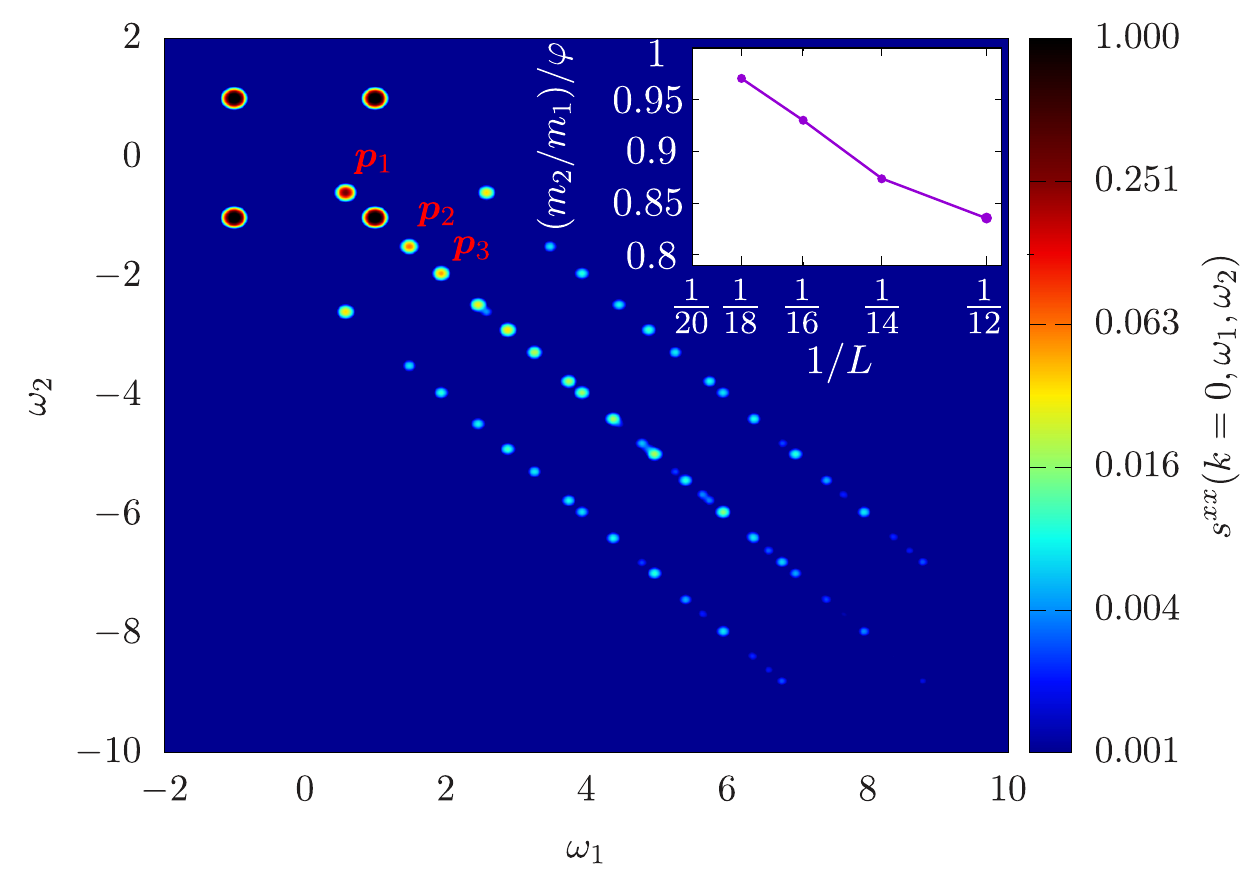}
\caption{Dynamical correlation function $s^{xx}(k=0,\omega_1,\omega_2)$ (\ref{eq:HMesons}) for the perturbed critical model with $J = 1$ ,$\Omega = 1$,  $h_z = \Omega/2+0.1$, $g = 1.5$ and $L = 16$. The data was normalized so that the maximum value of the plot is one. The four dominant peaks at $(\pm \Omega,\pm \Omega)$ and $(\mp \Omega,\pm \Omega)$ come from the driving frequency $\Omega$, while the next three  at $\bm p_i=(m_i-\Omega,\Omega-m_i)$ come from the masses of the mesons excitations $m_i$. Although $m_2/m_1$ is not equal to the golden ratio $\varphi$, we verify at the inset that, as  the size of the system is increased, $m_2/m_1$ gets closer to $\varphi$.}
\label{fig:pertcrit}
\end{figure}

\textit{Realizing a perturbed critical model.---}Let us finish with an illustration of the second interesting case discussed above. We consider tuning the driving such that the static Hamiltonian describes the perturbed critical Ising chain:
\begin{equation}
  H^1_{\rm st} = -J \sum_{l=1}^{L} \sigma^z_l \sigma^z_{l+1} - J \sum_{l=1}^L \sigma^x_l  + h \sum_{l=1}^L \sigma^z_l\,.
  \label{eq:HMesons} 
\end{equation}
When $h=0$, $H^1_{\rm st}$ realizes the critical point of the Ising chain. For $h\neq0$, $H^1_{\rm st}$ is no longer integrable, but its low-energy physics is well-understood thanks to Zamolodchikov~\cite{zamolodchikov1989integrals}. Pairs of fermions (corresponding to domain walls in the ordered phase) are confined by the presence of the longitudinal field $h$ and form ``meson'' excitations. In the scaling limit, the algebraic structure of the theory allows the prediction of these meson masses, including the beautiful result that the ratio of masses of the first and second meson states realizes the golden ratio, $m_2/m_1 = \varphi$. With integrability absent, we are limit to performing small system numerics, such as in Fig.~\ref{fig:pertcrit}.

In Fig.~\ref{fig:pertcrit} we plot the dynamical correlation function
\begin{equation}
\begin{split}
s^{xx}(k=0,\omega_1,\omega_2)&=\sum_{\ell}\int dt_1 dt_2 
		 e^{i (\omega_1 t_1+\omega_2 t_2)}
					   \\& \times\langle \Psi_0 | \sigma^x_{j+\ell}(t_1) \sigma^x_{j}(t_2) | \Psi_0 \rangle,
\end{split}
\label{eq:dynamicCorr} 
\end{equation}
where $\sigma^x_{n}(t)$ denotes the time-evolution of $\sigma^x_{n}$ in the Heisenberg picture and, for simplicity, we assume the initial state of the system was prepared to be the ground state of the static Hamiltonian (\ref{eq:HMesons}). Note that, because of the driving, energy is not conserved, and $\langle \Psi_0 | \sigma^x_{j+\ell}(t_1) \sigma^x_{j}(t_2) | \Psi_0 \rangle$ is no longer a function of the time difference $(t_1-t_2)$, therefore we considered the Fourier transform of both times.

Note that there are  four dominant peaks in Fig.~\ref{fig:pertcrit}, these correspond to the driving frequency $\Omega$ and are located at $(\omega_1,\omega_2)=(\pm \Omega,\pm \Omega),(\mp \Omega,\pm \Omega)$.  The remaining dominant peaks (marked as $\bm p_1$, $\bm p_2$ and $\bm p_3$ in the figure) correspond to the first excitations or  masses of the static system, and their coordinates are $\bm p_i=(m_i-\Omega,\Omega-m_i)$, where $m_i$ denotes the masses of the meson excitations. Although these masses do not satisfy the equality $m_2/m_1=\varphi$, we verify that this is a finite size effect in the inset of the figure, as $m_2/m_1$ gets closer to $\varphi$ when the size of the system is increased.  This reiterates the fact that driven systems, such as~\eqref{Ht}, can realise exotic physics that is inaccessible in its undriven state.

\textit{Discussion.---}In this Letter, we have explored an example of driven system that is instantaneously nonintegrable but can nonetheless be solved exactly. This is due to a hidden integrability in the problem, that is not apparent from the time-dependent Schr\"odinger equation: the instantaneous Hamiltonian  $H(t)$ is nonintegrable, but dynamics of observables are nonetheless controlled by an effective static, integrable Hamiltonian. This may provide a route to protecting quantum information from the scrambling associated with thermalization through the addition of driving; this is an interesting direction for future studies. 

The methods applied within this Letter can be used to tackle the dynamics of an infinite family of Hamiltonians (not necessarily integrable). This family of driven systems does not undergo heating to infinite temperature, even though they are absent disorder and (generically) integrability.  For example, consider the Hamiltonian $\widetilde H$ of any spin-1/2 chain that conserves total $\sigma^z_\text{tot}$ magnetization (this need not be translationally invariant), which is driven as in~\eqref{Ht}:
\begin{equation}
  \widetilde H(t) = \widetilde H - \widetilde g \sum_l \left( e^{-i\widetilde\Omega t} \sigma^+_l + \text{H.c.}\right).
\end{equation}
The transformation~\eqref{Ut} still maps Eq.~\eqref{SchrodingerEq} to a time-independent Schr\"odinger equation with the new effective static Hamiltonian $\widetilde H_\text{st} = \widetilde H-(\widetilde\Omega/2) \sum_l \sigma^z_l - \widetilde g \sum_l \sigma^x_l$.  Time-evolution of observables in a driven system has once again been mapped to a sudden quench problem. It would be interesting to explore this idea further in interacting models, such as when $\tilde H_{\text{st}}$ describes the Heisenberg or XXZ model, where potentially integrability can be harnessed to perform exact calculations. 

Another scenario worthy of attention is to consider a problem in which the parameters of the static Hamiltonian describe a different phase to the initial Hamiltonian. One may then expect to see signatures of dynamical phase transitions in the nonequilibrium dynamics, such as kinks in the Lochsmidt echo~\cite{heyl2013dynamical,karrasch2013dynamical}. Further exploring the lattice limit of Zamolodchikov's perturbed Ising field theory~\cite{zamolodchikov1989integrals}, which features interesting collective excitations related to an exotic hidden $E_8$ algebraic structure, is interesting. Such studies would require detailed numerical analysis (perhaps in the scaling limit~\cite{james2018nonperturbative}), an avenue left to future works.

\begin{acknowledgments}
\textit{Acknowledgments.---} Our thanks go to Jean-S\'ebastien Caux, Mario Collura, Andrew James, Robert Konik, Tam\'as P\'almai, and Sergio Tapias Arze for useful discussions. I.P.C. and E.G.-G.  thanks the London Mathematical Laboratory for financial support. N.J.R. was supported by funding from the EU's Horizon 2020 research and innovation programme, under grant agreement No~745944, and the European Research Council under ERC Advanced Grant No~743032 ({\sc Dynamint}). E.G.-G. acknowledges the hospitality of the Institute of Physics, University of Amsterdam, during the completion of this work. N.J.R. thanks the Institute of Physics of the  National Autonomous University of Mexico for hospitality during a visit, where part of this work was undertaken. 
\end{acknowledgments}

\bibliography{ising.bib}

\setcounter{equation}{0}
\setcounter{figure}{0}
\setcounter{table}{0}
\makeatletter
\renewcommand{\theequation}{S\arabic{equation}}
\renewcommand{\thefigure}{S\arabic{figure}}
\renewcommand{\thesection}{S\arabic{section}}

\onecolumngrid

\newpage
\begin{center}
  {\large\textbf{Supplemental Material for: ``Quantum quench in a driven Ising chain''}}\\
  \vspace{2mm}
  Neil J. Robinson$^{1}$, Isaac P\'erez Castillo,$^{2}$ and Edgar Guzm\'an-Gonz\'alez$^{3}$\\
  \vspace{2mm}
  \textit{\small$^{1}$Institute for Theoretical Physics, University of Amsterdam,\\
    Postbus 94485, 1090 GL Amsterdam, The Netherlands\\
     \vspace{1mm}$^{2}$Departamento de F\'isica, Universidad Autónoma Metropolitana-Iztapalapa, \\San Rafael Atlixco 186, Ciudad de México 09340, Mexico\\
    \vspace{1mm}$^{3}$London Mathematical Laboratory, 18 Margravine Gardens, London W6 8RH, United Kingdom}\\
   
\end{center}

\twocolumngrid

In this Supplemental Material, we present: 
\begin{enumerate}
\item The Magnus expansion for the problem in the main text.
\item The exact mapping between time-dependent two-point functions of the driven systems and a sudden quench problem.
\item Exact full-time evolution of one- and two-point functions for a special case of the driven system quench. 
\item Details of the computation of ``sudden quench'' correlation functions, summarizing both previously known results and presenting new ones.
\item Details of our numerical algorithm for simulating time-evolution in the driven system.
\item Details of the procedure used to compare finite-size numerical results with analytical expressions.
\end{enumerate}

\section{Magnus Expansion} 

An alternative approach to compute the time-evolution of the driven system is given by the Magnus expansion. For a time-dependent Hamiltonian $H(t)$, the time-evolution operator ${\cal U}(t)$ is defined via the differential equation
\begin{equation}
  i\frac{\rd}{\rd t}{\cal U}(t) = H(t){\cal U}(t), \qquad {\cal U}(t=0) = \mathbb{1}. 
\end{equation}
This is formally solved in terms of the well-known time-ordered exponential
\begin{equation}
  {\cal U}(t) = \mathbb{T}\,\exp\left[ -i\int_0^t \rd t' H(t')\right].
\end{equation}
However, the time-ordering here (denoted by the operator $\mathbb{T}$) is difficult to treat and so it may be beneficial to take an alternative approach. One such approximate approach is the Magnus expansion~[68], where the time-evolution operator is instead approximated by
\begin{equation}
  {\cal U}(t) = \exp\left[ \Xi(t)\right], \qquad \Xi(t) = \sum_{k=1}^\infty \Xi_k(t),
\end{equation}
where terms in the series are obtained from the differential equation
\begin{equation}
  i\frac{\rd}{\rd t} \Xi(t) = \frac{\text{ad}_{\Xi}}{\exp(\text{ad}_\Xi)-1}\, H(t).
\end{equation}
Here $\text{ad}_{\Xi}$ is the adjoint action of $\Xi$: $\text{ad}_{\Xi}(O) = [\Xi,O]$. The terms of the series $\Xi_k(t)$ have well-known forms
\begin{widetext}
\begin{equation}
  \begin{split}
    \Xi_1(t) =& -i\int_0^t \rd t_1 H(t_1),\qquad
    \Xi_2(t) = -\frac{1}{2}\int_0^t\rd t_1\int_0^{t_1}\rd t_2\left[ H(t_1), H(t_2) \right], \\
    \Xi_3(t) =& \frac{i}{6} \int_0^{t} \rd t_1 \int_0^{t_1} \rd t_2 \int_0^{t_2} \rd t_3 \Big( \big[H(t_1),[ H(t_2), H(t_3)]\big] + \big[H(t_3),[H(t_2),H(t_1)]\big]\Big).  
  \end{split}
\end{equation}
\end{widetext}
Here the first term simply corresponds to taking the time-averaged Hamiltonian, as might well be expected.

Let us now turn to the problem at hand in the main body of the manuscript, and examine the Magnus expansion for the time-evolution operator, working at small $T$, i.e. high frequency for the driving. We will see that the integrability that we harnessed in the main text to perform exact calculations is completely hidden in the Magnus expansion. Indeed, the presence of a simple static limit (independent of integrability of that static Hamiltonian) is not at all apparent. To illustrate this, we consider the time-evolution operator over a single period $T=2\pi/\Omega$ of the driving, ${\cal U}(T)$. Treating the Magnus expansion as a power series approximation in $T$, the first order term in the expansion is just the average of the Hamiltonian over a single period, so the driving vanishes:
\begin{equation}
  i \Xi_1(T) = -iT \left( -J \sum_{l=1}^{L} \sigma^z_l \sigma^z_{l+1} + h^z \sum_{l=1}^{L} \sigma^z_l\right). 
\end{equation}

For the second order term in the expansion, we require the commutator
\begin{equation}
  \begin{split}
    [H(t_1),H(t_2)] =&\ 2ig\sum_{l=1}^L  \Bigg\{  g \sin\Big(\Omega(t_2-t_1)\Big)\sigma^z_l \nonumber\\
    &- h^z \Big[ (c_2-c_1)  \sigma^y_l -  (s_2-s_1) \sigma^x_l \Big] \\
    & \hspace{0cm} + J (c_2 - c_1) \Big( \sigma^z_l \sigma^y_{l+1} + \sigma^y_l \sigma^z_{l+1}\Big)\\
    &-J(s_2 - s_1) \Big( \sigma^z_l \sigma^x_{l+1} + \sigma^x_l \sigma^z_{l+1}\Big)\Bigg\}.  
  \end{split}
\end{equation}
where for conciseness we define the short hand notations  $c_a = \cos(\Omega t_a)$, $s_a = \sin(\Omega t_a)$. Performing the double integral, we obtain $\Xi_2(T)$:
\begin{equation}
	i\Xi_2(T)= \frac{gT^2}{\pi} \sum_{l=1}^L \bigg[ h^z\sigma^x_l -\frac{g}{2} \sigma^z_l - J \Big(\sigma^z_l\sigma^x_{l+1}+\sigma^x_l\sigma^z_{l+1}\Big)\bigg].
\end{equation}
In this second order term (in $T$), we see already that the Magnus expansion for the time-evolution over one period is rather messy. Continuing to third order, we expect to generate terms of the form ${\sigma^x_l \sigma^y_{l+1},\, \sigma^y_l \sigma^z_{l+1}}$ too, so higher orders are unlikely to simplify things further. We see that the existence of a simple static Hamiltonian, as given in the main body of the manuscript, is not obvious from the Magnus expansion.

\section{Mapping between driven and ``sudden quench''  two-point functions}
Let us consider equal-time two-point functions in the nonequilibrium time-evolved wave function [see Eq.~(7) of the main text],
\begin{equation}
  s^{\alpha\beta}(\ell;t) = \la \Psi(t)|\sigma^\alpha_j \sigma^\beta_{j+\ell} |\Psi(t)\ra.
\end{equation}
Using that $U(t)$ is an element of SU(2) leads to the transformations described in Eq.~(8) of the main text, and we arrive at the following mapping between driven and ``sudden quench'' correlation functions:
\begin{equation}
  \begin{split}
    s^{zz}(\ell;t) =&\ s^{zz}_{\text{st}}(\ell;t), \\
    s^{xx}(\ell;t) =&\ \cos^2(\Omega t)s^{xx}_{\text{st}}(\ell;t) + \sin^2(\Omega t) s^{yy}_{\text{st}}(\ell;t)\\ 
    &- \frac12 \sin(2\Omega t) \big[ s^{xy}_{\text{st}}(\ell;t) + s^{yx}_{\text{st}}(\ell;t)\big], \\
    s^{yy}(\ell;t) =&\  \cos^2(\Omega t) s^{yy}_{\text{st}}(\ell;t) + \sin^2(\Omega t) s^{xx}_{\text{st}}(\ell;t)\\
    &+ \frac12 \sin(2\Omega t) \big[ s^{xy}_{\text{st}}(\ell;t) + s^{yx}_{\text{st}}(\ell;t)\big], \\
    s^{xy}(\ell;t) =&\ \cos^2(\Omega t) s^{xy}_{\text{st}}(\ell;t) - \sin^2(\Omega t) s^{yx}_{\text{st}}(\ell;t)\\
    &+ \frac12 \sin(2\Omega t) \big[ s^{xx}_{\text{st}}(\ell;t) - s^{yy}_{\text{st}}(\ell;t)\big],\\
    s^{xz}(\ell;t) =&\ \cos(\Omega t) s^{xz}_{\text{st}}(\ell;t) - \sin(\Omega t)s^{yz}_{\text{st}}(\ell;t), \\
    s^{yz}(\ell;t) =&\  \cos(\Omega t) s^{yz}_{\text{st}}(\ell;t) + \sin(\Omega t)s^{xz}_{\text{st}}(\ell;t).
  \end{split}
  \label{twopts}
\end{equation}
Here, as in the main body of the manuscript, we define the ``sudden quench'' correlation functions
\begin{align}
  s^{\alpha\beta}_{\text{st}}(\ell;t)= \sum_{n,m}&e^{i(E_n-E_m)t}\la E_n|\sigma^\alpha_j\sigma^\beta_{j+\ell}|E_m\ra \nonumber\\
                                              &\times \la\Psi_0|E_n\ra \la E_m |\Psi_0\ra,
  \label{salpbet}
\end{align}
where $|E_n\ra$ are eigenstates of the static Hamiltonian $H_\text{st}$ with energy $E_n$ and $|\Psi_0\ra = |\Psi(t=0)\ra$ is the initial state for the time-evolution. We note that Eq.~\eqref{salpbet} is the same as time-evolving the initial state with $H_\text{st}$ alone. 

\section{Full time-evolution of a special case}
In this section, we consider a slightly surprising scenario from our mapping. We consider starting in the ground state of the Hamiltonian
\begin{equation}
H_{\text{init}} = -J \sum_{l=1}^L \sigma^z_{l}\sigma^z_{l+1} - g \sum_l \sigma^x_{l}. 
\end{equation}
Let us work with $J>0$ and $g>0$. We perform a quench where we start to rotate the transverse field \textit{and apply an additional longitudinal field}
\begin{align}
  H(t>0) = &- J\sum_{l=1}^L \sigma^z_l \sigma^z_{l+1} + h^z \sum_{l=1}^{N} \sigma^z_l \nonumber\\
  &- \sum_{l=1}^L g \Big( e^{-i\Omega t} \sigma^+_l + e^{i\Omega t}\sigma_l^-\Big).  
\end{align}
with frequency $\Omega$. We choose $h^z = \Omega/2$, in which case the effective static Hamiltonian is
\begin{equation}
H_{\text{st}} = -J\sum_{l=1}^L \sigma^z_l \sigma^z_{l+1} - g \sum_{l=1}^L \sigma^x_l.
\end{equation}
That is, the effective state Hamiltonian is identical to the initial Hamiltonian! In this case, the ``sudden quench'' correlation functions reduce to the equilibrium ones. These can be computed with standard free fermion techniques, see Ref.~[43], and read:
\begin{align}
  s^z_{\text{st}}(\ell;t) &=  \left( 1 - \lambda^{-2}\right)^{1/8} \Theta(\lambda-1),  \\
  s^x_{\text{st}}(\ell;t) &=  G(0), \quad
  s^y_{\text{st}}(\ell;t) = 0,
\end{align}
for the one-point functions, while the diagonal two-point functions read:
\begin{equation}
  \begin{split}
    s^{zz}_{\text{st}}(\ell;t) &= \text{det} \left[
      \begin{array}{cccc}
        G(-1) & G(-2) & \ldots & G(-\ell)   \\
        G(0)  & G(-1) & \ldots & G(-\ell+1) \\
        \vdots&\vdots &        & \vdots  \\
        G(\ell-2) & G(\ell-1) & \ldots & G(-1)                        
      \end{array}\right]
        \\
    s^{xx}_{\text{st}}(\ell;t) &= - G(\ell)G(-\ell) +  G(0)^2, \\
    s^{yy}_{\text{st}}(\ell;t) &=  \text{det} \left[
      \begin{array}{cccc}
        G(1) & G(0) & \ldots & G(2-\ell)   \\
        G(2)  & G(1) & \ldots & G(3-\ell) \\
        \vdots&\vdots &        & \vdots  \\
        G(\ell) & G(\ell-1) & \ldots & G(1)                        
      \end{array}\right]
  \end{split}
\end{equation}
where we use notations, $\lambda = J/g$, $\Theta(x\leq0) = 0$, $\Theta(x>0) = 1$, and we define the functions
\begin{equation}
  G(n) = L(n) + \lambda L(n+1),
\end{equation}
with
\begin{equation}
    L(n) = \frac{1}{\pi} \int_0^\pi \rd k\, \frac{\cos(kn)}{\sqrt{1+\lambda^2+2\lambda\cos(k)}}. 
\end{equation}

The off-diagonal (in spin indices) two-point functions require some further work, not usually being considered in equilibrium treatments of the quantum Ising chain. By symmetry arguments $s^{\alpha\beta}(\ell,t)=s^{\beta\alpha}(\ell,t)$. Using the Heisenberg equation of motion, $s^{zy}_{\text{st}} (\ell,t)\propto \dot s^{zz}_{\text{st}}(\ell,t)=0$, as we are working in an static case. The terms $s^{xz}_{\text{st}}(\ell,t)$ and  $s^{xy}_{\text{st}}(\ell,t)$ can not be computed in general using the standard techniques, as they mix the even sector with the odd sector in the fermionic theory (see the next section for more details). However, in the paramagnetic phase, where $\lambda<1$, using the symmetry of the Hamiltonian under $\pi$ rotations of the quantization axis about the $x$-direction, after some tedious algebra, we find
\begin{align}
    s^{zy}_{\text{st}}(\ell;t) & = 0, \\
    \left. s^{xz}_{\text{st}}(\ell;t)\right |_{\lambda <1} &= 0, \\
    \left. s^{xy}_{\text{st}}(\ell;t)\right |_{\lambda <1} &=0 ,
\end{align}

Putting these known equilibrium results together with the mapping between driven correlation functions and ``sudden quench'' ones, we obtain the \textit{full time-dependent correlation functions} for the $H_\text{init}\to H(t>0)$ quench
\begin{equation}
  \begin{split}
    s^z(\ell;t) &= s^z_{\text{st}}(\ell;t) \\
    s^x(\ell;t) &= \cos(\Omega t) s^{x}_{\text{st}}(\ell;t) , \\
    s^y(\ell;t) &= \sin(\Omega t) s^x_{\text{st}}(\ell;t), \\
    s^{zz}(\ell;t) &= s^{zz}_{\text{st}}(\ell;t), \\
    s^{xx}(\ell;t)|_{\lambda <1} &= \cos^2(\Omega t)s^{xx}_{\text{st}}(\ell;t) + \sin^2(\Omega t)s^{yy}_{\text{st}}(\ell;t), \\
    s^{yy}(\ell;t)|_{\lambda <1} &= \cos^2(\Omega t)s^{yy}_{\text{st}}(\ell;t) + \sin^2(\Omega t)s^{xx}_{\text{st}}(\ell;t), \\
    s^{xy}(\ell;t)|_{\lambda <1} &= \sin(\Omega t)\cos(\Omega t) \Big[ s^{xx}_{\text{st}}(\ell;t) - s^{yy}_{\text{st}}(\ell;t)\Big], \\
    s^{xz}(\ell;t)|_{\lambda <1} &= 0 \\
    s^{yz}(\ell;t)|_{\lambda <1} &= 0\,.
  \end{split}
\end{equation}

Furthermore, in the limit $\ell\gg1$ one can simplify some of these results using an  asymptotic analysis via Szeg\"o's theorem, which leads to (some) closed form analytical results~[52--54]
\begin{widetext}
\begin{equation}
  \begin{split}
    s^{zz}_{\text{st}}(\ell\gg1;t)\bigg\vert_{\lambda < 1} \approx& \frac{1}{\sqrt{\pi}} \left( 1 - \lambda^2\right)^{-1/4} \frac{\lambda^{\ell}}{\sqrt{\ell}} \left[ 1 - \frac{1}{8\ell} \frac{1+\lambda^2}{1-\lambda^2} + O\left(\frac{1}{\ell^{2}}\right)\right],\\
    s^{zz}_{\text{st}}(\ell\gg1;t)\bigg\vert_{\lambda > 1} \approx& \left( 1- \lambda^{-2}\right)^{1/4} + O\left(\frac{1}{\ell}\right),\\
    s^{xx}_{\text{st}}(\ell\gg1;t)\bigg\vert_{\lambda<1} \approx&
	\left(s^{x}_{\text{st}}(\ell;t)\right)^2
    -\frac{\lambda^{2\ell}}{2\ell^2 \pi}\left(1-\frac{1}{\ell}\frac{3\lambda^{-2}-1}{\lambda^{-2}+1}\right)
								   \\
    s^{xx}_{\text{st}}(\ell\gg1;t)\bigg\vert_{\lambda>1} \approx& 
    \left(s^{x}_{\text{st}}(\ell;t)\right)^2-\frac{1}{2\lambda^{2\ell+2}\pi}\left(1-\frac{1}{\ell}\frac{\lambda^{-2}-3}{\lambda^{-2}-1}\right)
    \\
    s^{yy}_{\text{st}}(\ell\gg1;t)\bigg\vert_{\lambda < 1} \approx& 
    -\frac{1}{2\sqrt{\pi}} (1-\lambda^2)^{3/4} \frac{\lambda^\ell}{\ell^{3/2}} \left[1 + \frac{3}{2\ell} \left(\frac{1}{2(1-\lambda^2)}- \frac14 \right) + O\left(\frac{1}{\ell^2}\right) \right] \\
    s^{yy}_{\text{st}}(\ell\gg1;t)\bigg\vert_{\lambda > 1} \approx&-\frac{1}{\pi} \left(1-\lambda^{-2}\right)^{-3/4} \frac{\lambda^{-2\ell}}{\ell^3} \left[\frac{1}{2} - \frac3{8\ell} \frac{1+\lambda^{-2}}{1-\lambda^{-2}}  + O\left(\frac{1}{\ell^{2}}\right)\right].\\
  \end{split}  
\end{equation}

Putting together these results, we obtain exact, closed form results for one-point functions and the asymptotics of two-point correlation functions in a nonequilibrium driven problem. The non-equilibrium one-point functions read:
\begin{align}
  s^{z}(t) &= \left(1-\lambda^{-2}\right)^{1/8} \Theta(\lambda-1), \\
  s^{x}(t)\Big\vert_{\lambda\neq1} &= \frac{\cos(\Omega t)}{\pi|1-\lambda|}\Bigg[
               \left(\lambda-1\right)^2 {\cal E}\left(- \frac{4\lambda}{(1-\lambda)^2} \right)
               -\left(\lambda^2-1\right) {\cal K}\left(- \frac{4\lambda}{(1-\lambda)^2} \right)
               \Bigg],\\
  s^{y}(t)\Big\vert_{\lambda\neq1} &= \frac{\sin(\Omega t)}{\pi|1-\lambda|}\Bigg[
               \left(\lambda-1\right)^2 {\cal E}\left(- \frac{4\lambda}{(1-\lambda)^2} \right)
               -\left(\lambda^2 -1\right){\cal K}\left(- \frac{4\lambda}{(1-\lambda)^2} \right)
               \Bigg]
\end{align}
where ${\cal E}(m) = \int_0^{\pi/2} \mathrm{d}\theta\, [1-m \sin^2(\theta)]^{1/2}$ and ${\cal K}(m) = \int_0^{\pi/2} \mathrm{d}\theta\, [1-m \sin^2(\theta)]^{-1/2} $ are elliptic functions. The asymptotics of non-equilibrium two-point functions read:
\begin{align}
  s^{zz}(\ell\gg1;t)\bigg\vert_{\lambda < 1}
  =&~ \frac{1}{\sqrt{\pi}} \left( 1 - \lambda^2\right)^{-1/4} \frac{\lambda^{\ell}}{\sqrt{\ell}}
     \left[ 1 - \frac{1}{8\ell} \frac{1+\lambda^2}{1-\lambda^2} + O\left(\frac{1}{\ell^{2}}\right)\right],\\
  s^{zz}(\ell\gg1;t)\bigg\vert_{\lambda > 1}
  =&~ \left( 1- \lambda^{-2}\right)^{1/4} + O\left(\frac{1}{\ell}\right), \\
  s^{xx}(\ell\gg1;t)\bigg\vert_{\lambda < 1}
  =&~
     \cos^2(\Omega t) \Bigg[
     \left(s^{x}_{\text{st}}(\ell;t)\right)^2
     -\frac{\lambda^{2\ell}}{2\ell^2\pi}
     \left(1-\frac{1}{\ell}\frac{3-\lambda^2}{1+\lambda^2}\right)
     \Bigg] \nonumber\\
   &~ -\sin^2(\Omega t) \Bigg[
     \frac{1}{2\sqrt{\pi}} (1-\lambda^2)^{3/4} \frac{\lambda^\ell}{\ell^{3/2}}
     \left[1 + \frac{3}{2\ell} \left(\frac{1}{2(1-\lambda^2)}
     -\frac14 \right) + O\left(\frac{1}{\ell^2}\right) \right] \\
  s^{yy}(\ell;t)\bigg\vert_{\lambda < 1} =&~
     \sin^2(\Omega t) \Bigg[
     \left(s^{x}_{\text{st}}(\ell;t)\right)^2
     -\frac{\lambda^{2\ell}}{2\ell^2\pi}
     \left(1-\frac{1}{\ell}\frac{3-\lambda^2}{1+\lambda^2}\right)
     \Bigg] \nonumber\\
   &~ -\cos^2(\Omega t) \Bigg[
     \frac{1}{2\sqrt{\pi}} (1-\lambda^2)^{3/4} \frac{\lambda^\ell}{\ell^{3/2}}
     \left[1 + \frac{3}{2\ell} \left(\frac{1}{2(1-\lambda^2)}
     -\frac14 \right) + O\left(\frac{1}{\ell^2}\right) \right]\\
  s^{xy}(\ell;t)\bigg\vert_{\lambda < 1} =&~
     \sin(\Omega t)\cos(\Omega t) \Bigg[
     \left(s^{x}_{\text{st}}(\ell;t)\right)^2
     -\frac{\lambda^{2\ell}}{2\ell^2\pi}
     \left(1-\frac{1}{\ell}\frac{3-\lambda^2}{1+\lambda^2}\right)
      \nonumber\\
   &~ + 
     \frac{1}{2\sqrt{\pi}} (1-\lambda^2)^{3/4} \frac{\lambda^\ell}{\ell^{3/2}}
     \left[1 + \frac{3}{2\ell} \left(\frac{1}{2(1-\lambda^2)}
     -\frac14 \right) + O\left(\frac{1}{\ell^2}\right) \right]
     \Bigg]\\
  s^{xz}(\ell;t)\bigg\vert_{\lambda < 1} =&~0 \\
  s^{yz}(\ell;t)\bigg\vert_{\lambda < 1} =&~0\,.
\end{align}
\end{widetext}

\section{The ``sudden quench'' correlation functions}
In this appendix we detail calculations of the ``sudden quench'' correlation functions. We follow the approach of Calabrese, Essler and Fagotti~[49--51] in computing one- and two-point correlation functions following a quantum quench of the transverse field in the quantum Ising chain. Much of the discussion will follow similar lines, although we will consider a number of observables that were not computed in their work.

\subsection{Fermionization of the model}
Let us begin by describing the exact solution of the transverse field Ising model in terms of free fermions. The transverse field Ising model has the Hamiltonian 
\begin{equation}
H = -J \sum_{\eel=1}^L \sigma^z_\eel \sigma^z_{\eel+1} - g \sum_{\eel=1}^L \sigma^x_\eel\,, 
\end{equation}
where {$\sigma^\alpha_\eel$} are the spin-1/2 operators acting on the $\eel$th site of an $L$ chain (we take $L$ even), $J$ is the exchange interaction, and $g$ is the transverse field strength. We consider periodic boundary conditions, identifying $\sigma^\alpha_{L+1} = \sigma^\alpha_1$. To express the model in terms of free fermions, 
we first perform a $3\pi/2$ rotation of the quantization axes about the $y$ axis to be consistent with standard notations
\begin{align}
  H &= - J\sum_{\eel=1}^L \tau^x_\eel \tau^x_{\eel+1} - g \sum_{\eel=1}^L \tau^z_\eel \nonumber\\
  &\equiv - \bar J \sum_{\eel=1}^L \Bigg( \tau^x_\eel \tau^x_{\eel+1} + \bar g \tau^z_{\eel} \Bigg). 
\end{align}
Here $\tau^\alpha_\eel$ are the  Pauli matrices with the rotated quantization axes, $\bar J = J$ and $\bar g = g/J$.

We proceed to fermionize the problem via the Jordan-Wigner transformation~[69]
\begin{equation}
  \begin{split}
  \tau^z_\eel &= 1 - 2a^\dagger_\eel a_\eel, \\
  \tau^-_\eel &= \exp\left( i\pi \sum_{j<\eel} a^\dagger_j a_j \right) a^\dagger_\eel, \\
  \tau^+_\eel &= a_\eel\exp\left(i\pi \sum_{j<\eel} a^\dagger_ja_j\right).
\end{split}
\label{jordanWigner}
\end{equation}
where we define the raising/lowering operators $\tau^\pm = (\tau^x \pm i\tau^y)/2$ and the fermions have canonical anticommutation relations $\{a_\eel, a^\dagger_{\eel'}\} = \delta_{\eel,\eel'}$. Following this transformation, the Hamiltonian reads
\begin{equation}
  \begin{split}
    H =& -\bar J \sum_{\eel=1}^{L-1} \Big(a^\dagger_\eel - a_\eel\Big)\Big(a_{\eel+1}+a_{\eel+1}^\dagger\Big)\\
    &+ \bar Je^{i\pi\hat N} \Big(a_L - a_L^\dagger\Big)\Big(a_1 + a_1^\dagger\Big)\\
    &+ \bar J \bar g \sum_{\eel=1}^{L}\Big(a^\dagger_\eel a_\eel - a_\eel a^\dagger_\eel\Big),
  \end{split}
  \label{jordanWignerH}
\end{equation}
where $\hat N = \sum_{\eel=1}^{L} a^\dagger_\eel a_\eel$ is the total number operator for the fermions. It is easy to see that fermion number parity is conserved by the Hamiltonian, so the exponential in the second term is
\begin{equation}
  \exp(i\pi\hat N) = \left\{ \begin{array}{lcl}
                               +1 & \quad & \text{for even numbers of fermions}, \\
                               -1 & \quad & \text{for odd numbers of fermions}.
                             \end{array}
                           \right.
\end{equation}
Thus the Hilbert space ${\cal H} = {\cal H}_e \otimes {\cal H}_o$ of the model splits into two sectors, for $\la\hat N\ra$ even or odd, which correspond to antiperiodic or periodic boundary conditions on the fermions. We need to address each of these cases in turn.

\subsubsection{Even numbers of fermions}
In the even sector, we have $\exp(i\pi \hat N) = +1$ and the Hamiltonian~\eqref{jordanWignerH} reads
\begin{equation}
  \begin{split}
  H_e =& -\bar J \sum_{\eel=1}^{L} \Big(a^\dagger_\eel - a_\eel\Big)\Big(a_{\eel+1}+a_{\eel+1}^\dagger\Big)\\
  &+ \bar J \bar g \sum_{\eel=1}^{L}\Big(a^\dagger_\eel a_\eel - a_\eel a^\dagger_\eel\Big),
  \end{split}
\end{equation}
where the fermions obey \textit{antiperiodic boundary conditions} $a_{L+1} = -a_1$. The Hamiltonian can be diagonalized by first Fourier transforming and then performing a Bogoliubov rotation. The Fourier transform reads
\begin{equation}
  c_{k_n} = \frac{1}{\sqrt{L}} \sum_{\eel=1}^L e^{ik_n\eel} a_\eel, 
\end{equation}
with $k_n = \pi(2n+1)/L$ with $n=-L/2,-L/2+1,\ldots,L/2-1$. In terms of the Fourier modes, the Hamiltonian reads
\begin{equation}
  \begin{split}
    H_e = -\bar J \sum_n &\bigg[ \cos(k_n) \left( c^\dagger_{k_n}c_{k_n} - c_{-k_n}c_{-k_n}^\dagger \right)\\
    &- i \sin(k_n) \left( c^\dagger_{k_n} c^\dagger_{-k_n} - c_{-k_n}c_{k_n}\right)\\
    &- \bar g \left( c^\dagger_{k_n}c_{k_n} - c_{-k_n}c_{-k_n}^\dagger\right) \bigg].
    \end{split}
\end{equation}
This can be diagonalized via the Bogoliubov rotation
\begin{equation}
  \left( \begin{array}{c} c_{k_n} \\ c^\dagger_{-k_n} \end{array}\right)
  =
  \left( \begin{array}{cc}
           \cos\left(\frac{\theta_{k_n}}{2}\right) & i\sin\left(\frac{\theta_{k_n}}{2}\right) \\
           i\sin\left(\frac{\theta_{k_n}}{2}\right) & \cos\left(\frac{\theta_{k_n}}{2}\right)
         \end{array}
  \right)
  \left(
    \begin{array}{c} \alpha_{k_n} \\ \alpha^\dagger_{-k_n} \end{array}\right),
  \label{bogoliubov}
\end{equation}
where the Bogoliubov angle $\theta_{k_n}$ satisfies
\begin{equation}
  \theta_{k_n}= -\text{sgn}(k_n) \arccos \left( \frac{\bar g-\cos(k_n)}{\sqrt{\bar g^2 - 2\bar g \cos(k_n) + 1}}\right)
\end{equation}     
to obtain the Hamiltonian
\begin{equation}
H_e = \sum_n \epsilon_{\bar g}(k_n) \Big( \alpha^\dagger_{k_n} \alpha_{k_n} - \frac12\Big), 
\end{equation}
with dispersion relation
\begin{equation}
  \epsilon_{\bar g}(k_n) = 2\bar J \sqrt{ \bar g^2 + 2\bar g \cos(k_n) + 1}.
\end{equation}
Thus the Hamiltonian has been diagonalized.

Let us now say a few words about the nature of the ground state in the even sector. As is expected from a Bogoliubov rotation (and as we well know from physics of the Ising chain), the dispersion is generically gapped, i.e. $\epsilon_{\bar g}(k_n) > 0$. As a result, the ground state $|\Omega_e,\bar g\ra$ is the vacuum for the Bogoliubov fermions:
\begin{equation}
\alpha_{k_n} |\Omega_e,\bar g\ra = 0, \quad \forall n. \label{evenGS}
\end{equation}
It immediately follows that the ground state energy is
\begin{equation}
  E_{\Omega_e,\bar g} = - \frac12 \sum_n \epsilon_{\bar g}(k_n). \label{evenGSen}
\end{equation}

It will be useful to understand the ground state $|\Omega_e,\bar g\ra$ in terms of the Jordan-Wigner fermions $c_{k_n}$. Inverting Eq.~\eqref{bogoliubov} to express $\alpha_{k_n}$ in terms $c_{k_n}$, one finds that Eq.~\eqref{evenGS} becomes ($\forall n$)
\begin{equation}
  \Big[ \cos\left(\frac{\theta_{k_n}}{2}\right) c_{k_n} - i \sin\left(\frac{\theta_{k_n}}{2}\right) c^\dagger_{-k_n} \Big] |\Omega_e,\bar g\ra = 0.
\end{equation}
As $|\Omega_e,\bar g\ra$ is in the even sector with zero momentum, it must be of the form $\prod_{k_n\ge 0} (a + b c^\dagger_{-k_n}c^\dagger_{k_n})|0\ra$, where $|0\ra$ is the vacuum for Jordan-Wigner fermions. Working through some algebra, one finds
\begin{equation}
|\Omega_e,\bar g\ra = \prod_{k_n\ge0} \left[ \cos\left(\frac{\theta_{k_n}}{2}\right) - i\sin\left(\frac{\theta_{k_n}}{2}\right)c^\dagger_{-k_n}c_{k_n}^\dagger \right] |0\ra.  
\end{equation}
This can be written in a ``squeezed state'' form following some simple manipulations
\begin{equation}
  \begin{split}
  |\Omega_e,\bar g\ra =& \left[\prod_{k_n\ge0} \cos\left(\frac{\theta_{k_n}}{2}\right)  \right]\\
  &\times \exp\left[ -i \sum_{k_n \ge 0} \tan\left( \frac{\theta_{k_n}}{2}\right)c^\dagger_{-k_n}c^\dagger_{k_n}\right] |0\ra.
  \end{split}
\end{equation}

\subsubsection{Odd number of fermions}
The story with the odd fermion number sector is similar to the even one. This time we have $\exp(i\pi\hat N) = -1$ and instead of antiperiodic boundary conditions, we have periodic ones $a_{L+1} = a_1$. The Fourier transform of the Jordan-Wigner fermions then reads
\begin{equation}
  c_{p_n} = \frac{1}{\sqrt{L}} \sum_\eel e^{ip_n\eel}a_\eel
\end{equation}
with the momenta $p_n = 2n\pi/L$ and $n=-L/2,-L/2+1,\ldots...$. We again perform a Bogoliubov rotation
\begin{equation}
  \left( \begin{array}{c} c_{p_n} \\ c^\dagger_{-p_n} \end{array}\right)
  =
  \left( \begin{array}{cc}
           \cos\left(\frac{\theta_{p_n}}{2}\right) & i\sin\left(\frac{\theta_{p_n}}{2}\right) \\
           i\sin\left(\frac{\theta_{p_n}}{2}\right) & \cos\left(\frac{\theta_{p_n}}{2}\right)
         \end{array}
  \right)
  \left(
    \begin{array}{c} \alpha_{p_n} \\ \alpha^\dagger_{-p_n} \end{array}\right),   
\end{equation}
with the Bogoliubov angle as defined in the even sector, with the additional definition of $\theta_{p_0} = 0$. The Hamiltonian~\eqref{jordanWignerH} is then
\begin{equation}
H_o = \sum_{n\neq0} \epsilon_{\bar g}(p_n) \Big( \alpha^\dagger_{p_n} \alpha_{p_n} - \frac12\Big) - 2\bar J (1-\bar g) \Big( \alpha^\dagger_0 \alpha_0 - \frac12\Big).  
\end{equation}
We see now that the ground state in the odd sector is
\begin{equation}
  |\Omega_o\ra = \alpha^\dagger_0 |0_\alpha\ra, \label{oddGS}
\end{equation}
where $|0_\alpha\ra$ is the vacuum state for $\alpha_{p_n}$. The ground state $|\Omega_o\ra$ has energy
\begin{equation}
  E_{\Omega_o} = -\bar J ( 1 - \bar g) - \frac12 \sum_{n\neq0} \epsilon_{\bar g}(p_n).
\end{equation}
As with the even sector, it will be useful to know how to express the odd ground state, Eq.~\eqref{oddGS}, in terms of the Jordan-Wigner fermions. Following the same sequence of manipulations as the even sector, one can express the state in the form
\begin{equation}
  \begin{split}
    |\Omega_o,\bar g \ra =& \left( \prod_{p_n\ge0} \cos\left(\frac{\theta_{p_n}}{2}\right)\right)\\
    &\times \exp\left(-i \sum_{p_n \ge 0}\tan\left(\frac{\theta_{p_n}}{2}\right) c^\dagger_{-p_n}c^\dagger_{p_n}\right) c^\dagger_0 |0\ra.
  \end{split}   
\end{equation}
As expected, this state has \textit{odd} fermion parity. 

\subsection{Time-evolution following a quantum quench}
Let us now turn to the problem of describing time-evolution following a quantum quench. In particular, we consider the scenario where for time $t < 0$ the system is in an eigenstate of the Hamiltonian $H_i$ with $\bar g = g_i$ and at time $t=0$ this is suddenly changed to $\bar g = g_f$ and the state evolves under this new Hamiltonian $H_f$ for all times $t > 0$.

At times $t<0$ we can diagonalize the Hamiltonian in terms of fermions $\beta$ by performing the Bogoliubov rotation with angle $\phi$. For times $t>0$ we can similarly diagonalize the Hamiltonian in terms of fermions $\alpha$ via the Bogoliubov rotation with angle $\theta$. Equating the expressions for the Jordan-Wigner fermions $a_k$ in terms of the Bogoliubov fermions, we have
\begin{widetext}
\begin{align}
  \cos\left(\frac{\theta_{k_n}}{2}\right) \alpha_{k_n} + i \sin\left(\frac{\theta_{k_n}}{2}\right)\alpha_{-k_n}^\dagger
  &= \cos\left(\frac{\phi_{k_n}}{2}\right) \beta_{k_n} + i \sin\left(\frac{\phi_{k_n}}{2}\right) \beta^\dagger_{-k_n}, \\
  i\sin\left(\frac{\theta_{k_n}}{2}\right)\alpha_{k_n} + \cos\left(\frac{\theta_{k_n}}{2}\right) \alpha^\dagger_{-k_n}
  &= i \sin\left(\frac{\phi_{k_n}}{2}\right)\beta_{k_n} + \cos\left(\frac{\phi_{k_n}}{2}\right)\beta^\dagger_{-k_n}. 
\end{align}
The initial state, an eigenstate of $H_i$ will be expressed straightforwardly in terms of the $\beta$ fermions. Time-evolution, however, will be easy to compute in terms of the $\alpha$ fermions via
\begin{equation}
  \alpha_{k_n}(t) = \exp\left( - i\epsilon_{g_f}(k_n)t\right) \alpha_{k_n}.
\end{equation}
Thus we write the $\beta$ fermions in terms of $\alpha$ ones and time-evolve them
\begin{equation}
    \beta_{k_n}(t) = \cos\left(\frac{\theta_{k_n}-\phi_{k_n}}{2}\right) e^{-i\epsilon_{g_f}(k_n)t} \alpha_{k_n} + i \sin\left(\frac{\theta_{k_n}-\phi_{k_n}}{2}\right)e^{i\epsilon_{g_f}(k_n)t} \alpha^\dagger_{-k_n},
\end{equation}
before re-expressing the $\alpha$ fermions in terms of the $\beta$ ones:
\begin{equation}
\beta_{k_n}(t) = \bigg[ \cos\Big(\epsilon_{g_f}(k_n)t\Big) - i\sin\Big(\epsilon_{g_f}(k_n)t\Big)\cos\Big(\theta_{k_n}-\phi_{k_n}\Big)\bigg] \beta_{k_n} - \sin\Big(\epsilon_{g_f}(k_n)t\Big)\sin\Big(\theta_{k_n}-\phi_{k_n}\Big) \beta^\dagger_{-k_n}. \label{jwfermiont}
\end{equation}
\end{widetext}
We see that the difference in Bogoliubov angles often appears, so we denote this by the short hand
\begin{equation}
  \Delta_k = \theta_k - \phi_k.
\end{equation}
With this, and with some work, we can compute time-evolution of observables. In the following we first consider one-point functions, before continuing to two-point functions.

\subsection{Time-evolution of one-point functions} 

We will consider the quench where we start from the ground state of the initial Hamiltonian $H_i \equiv H(t<0)$ with $\bar g = g_i$ and time-evolve according to the Hamiltonian $H_f$ where $\bar g = g_f$. In the thermodynamic limit, the initial state will be a superposition of the ground states in the even and odd sectors
\begin{equation}
  |\Omega,\bar g\ra = \frac{1}{\sqrt{2}}\Big(|\Omega_e,\bar g\rra + |\Omega_o,\bar g\rra\Big).
  \label{gs}
\end{equation}
(Here we work with \textit{normalized} states such that $\lla \Omega_i, \bar g | \Omega_i, \bar g \rra = 1$.) Thus, generally, there will be three sorts of terms when computing observables: expectation values within the even sector, those within the odd sector, and those that mix the two sectors. Depending on the operator being considered, the mixing terms may be forbidden.

Let us begin with the easiest case, and consider the time-evolution of the one-point function of $\tau^z_\eel$. Under the Jordan-Wigner transformation~\eqref{jordanWigner} this maps to the local number operator of the Jordan-Wigner fermions. Thus we want to compute
\begin{equation}
 t^z(t) =  \la \Omega,g_i |e^{iH_ft}\Big( 1 - 2 a^\dagger_{\eel} a_{\eel} \Big) e^{-iH_ft} |\Omega,g_i\ra .
\end{equation}
Clearly $t^z(t)$ is independent of $\eel$ by translational invariance.
As $\tau^z$ is number conserving in terms of the fermions, the terms mixing even and odd sectors from Eq.~\eqref{gs} will vanish. Thus we have
\begin{equation}
  \begin{split}
  t^z(t) =&
  \lla\Omega_e,g_i|\bigg(\frac12-a^\dagger_{\eel}(t)a_{\eel}(t)\bigg) |\Omega_e,g_i\rra\\
  &+\lla\Omega_o,g_i|\bigg(\frac12-a^\dagger_{\eel}(t)a_{\eel}(t)\bigg) |\Omega_o,g_i\rra.
  \end{split}
\end{equation}
Let us take the first term, where the ground states is in the even sector of the Hilbert space. We express the time-evolved Jordan-Wigner fermions in terms of the time-evolved Bogoliubov fermions~\eqref{jwfermiont}. This yields
\begin{widetext}
\begin{align}
  a_\eel(t)_e = \frac{1}{\sqrt{L}}\sum_{n} e^{-ik_n\eel} c_{k_n}(t)
  = \frac{1}{\sqrt{L}}\sum_{n} e^{-ik_n\eel} \left[
  \cos\left(\frac{\phi_{k_n}}{2}\right)\beta_{k_n}(t)
  + i \sin\left(\frac{\phi_{k_n}}{2}\right) \beta^\dagger_{-k_n}(t) \right].
  \label{timeEvoJordanWigner}
\end{align}
Working through some tedious algebra, one finds
\begin{align}
  t^z(t)_e &= \frac12 - \frac{1}{L}\sum_n \left[ \sin^2\left(\epsilon_{g_f}(k_n)t\right)\sin^2\left(\Delta_{k_n} + \frac{\phi_{k_n}}{2}\right) + \sin^2\left(\frac{\phi_{k_n}}{2}\right)\cos^2\left(\epsilon_{g_f}(k_n)t\right)\right],\\
                &= \frac12 - \frac{1}{2\pi}\int_{-\pi}^{\pi} \rd k \left[ \sin^2\left(\epsilon_{g_f}(k)t\right)\sin^2\left(\Delta_{k} + \frac{\phi_{k}}{2}\right) + \sin^2\left(\frac{\phi_{k}}{2}\right)\cos^2\left(\epsilon_{g_f}(k)t\right)\right].
\end{align}
In the second line we use the thermodynamic limit to replace the sum over discrete momentum modes by an integral. Performing a similar calculation in the odd sector, we obtain the same result in the thermodynamic limit, leading to
\begin{equation}
  t^z(t) =  1 - \frac{1}{\pi}\int_{-\pi}^{\pi} \rd k \left[ \sin^2\left(\epsilon_{g_f}(k)t\right)\sin^2\left(\Delta_{k} + \frac{\phi_{k}}{2}\right) + \sin^2\left(\frac{\phi_{k}}{2}\right)\cos^2\left(\epsilon_{g_f}(k)t\right)\right].
  \label{timeEvoTz}
\end{equation}
\end{widetext}

Let us now move on to considering $\tau^x_l$ and $\tau^y_l$. These operators are similar in that they change the number of Jordan-Wigner fermions by one, and hence their expectation values consist only of the ``cross terms'' between even and odd ground states. This is actually quite problematic for the free fermion techniques discussed here, as it is not clear how to deal with such terms. Instead, we use the following trick: we relate the one-point function to the two-point function in the limit of infinite separation of the operators
\begin{equation}
\lim_{\ell\to\infty} \lla \Omega;g_i| \tau^x_{j}(t) \tau^x_{j+\ell}(t) |\Omega;g_i\rra \to \lla \Omega;g_i | \tau^x_j(t) |\Omega;g_i \rra^2,
\end{equation}
and similarly for $\tau^y$. As such, this naturally brings us on to the topic of computing two-point functions.

\subsection{Time-evolution of two-point functions}
We now turn our attention to computing two-point functions
\begin{equation}
  t^{\alpha\beta}(\ell;t) = \la \Omega, g_i | e^{iH_ft} \tau^\alpha_j \tau^\beta_{j+\ell} e^{-iH_f t} |\Omega, g_i \ra.
\end{equation}
Two examples of these, $\alpha=\beta=x$ and $\alpha=\beta=z$, were considered in Refs.~[49--51]. We will also consider these here, as well as some other cases that will be of use in this work. As with the one-point functions, we are unable to compute two-point functions when the operator within the expectation value is odd under spin inversion (and thus connects states in the even and odd sectors of the Hilbert space). Thus we restrict our attention to operators that are even under spin inversion.

This restriction, in the context of the main text, leads us to focus our results on one of two cases. Firstly, we can consider quenches starting in states that are even under spin inversion (and thus the expectation value vanishes by symmetry). This allows us, for example, to compute the full time evolution of two-point functions in the driven problem where the ``effective sudden quench'' is within the paramagnetic phase of the static Hamiltonian. Secondly, we can restrict attention to stroboscopic times, where contributions from these problematic expectation values vanish. 

\subsubsection{The $t^{zz}(\ell;t)$ two-point function} 
Let us begin with the easiest case, where the operators in the two-point function are solely in the transverse field direction. Then, according to the Jordan-Wigner transformation~\eqref{jordanWigner}, we need to compute the two-point function of the (time-dependent) Jordan-Wigner fermion density
\begin{widetext}
\begin{equation}
t^{zz}(\ell;t) = \la \Omega,g_i| e^{iH_ft} \Big( 1 -2a^\dagger_j a_j \Big)\Big( 1 - 2a^\dagger_{j+\ell}a_{j+\ell} \Big) e^{-iH_ft} |\Omega,g_i\ra. 
\end{equation}
The operator under consideration is even under spin inversion, thus matrix elements between states in the even and odd sectors vanish, leaving only
\begin{equation}
  t^{zz}(\ell;t) = \frac12 \la\la \Omega_e, g_i | e^{iH_ft} \Big( 1 - 2a^\dagger_j a_j \Big) \Big( 1 - 2a^\dagger_{j+\ell}a_{j+\ell} \Big) e^{-iH_ft} | \Omega_e , g_i \ra\ra + (e \leftrightarrow o).  
\end{equation}
Using Eq.~\eqref{timeEvoJordanWigner} (and similar for the odd sector) and taking the thermodynamic limit, one finds
\begin{align}
  t^{zz}(\ell;t) &= - \int_{-\pi}^{\pi} \frac{\rd k}{2\pi} \frac{\rd k'}{2\pi} \Bigg\{ e^{i(k-k')\ell} \Big[ \sin(2\epsilon_{g_f}(k)t) \sin(\Delta_k) \sin(2\epsilon_{g_f}(k')t)\sin(\Delta_{k'})   + \Big( e^{i(k-k')\ell} - 1 \Big) \nonumber \\
  & \qquad \times e^{i(\theta_{k}+\theta_{k'})} \Big(\cos(\Delta_k) - i \sin(\Delta_k) \cos(2\epsilon_{g_f}(k)t)\Big) \Big(\cos(\Delta_{k'}) - i \sin(\Delta_{k'}) \cos(2\epsilon_{g_f}(k')t)\Big) \Bigg\}.
  \label{timeEvoTzz}
\end{align}
\end{widetext}
\subsubsection{The $t^{xx}(\ell;t)$ two-point function}
Let us now turn our attention towards the two-point function where both operators are in the Ising direction. This is significantly more complicated than $t^{zz}(\ell;t)$ as the Jordan-Wigner transformation~\eqref{jordanWigner} introduces ``strings'' of density operators between $j$ and $j+\ell$. This can easily be seen by Taylor expanding the exponential factors in Eq.~\eqref{jordanWigner}:
\begin{widetext}
\begin{align}
  \exp\left( \pm i\pi \sum_{j<\ell} a^\dagger_j a_j \right) = \prod_{j < \ell} \exp\left( \pm i \pi a^\dagger_ja_j \right) = \prod_{j<\ell} \left( 1 - 2a_j^\dagger a_j\right). 
\end{align}
Thus, in terms of the Jordan-Wigner fermions, we need to compute
\begin{align}
  t^{xx}(\ell;t) &= \la \Omega,g_i | e^{iH_ft} (\tau^+_j + \tau^-_j ) (\tau^+_{j+\ell} + \tau^-_{j+\ell}) e^{-iH_ft} |\Omega,g_i \ra, \\
                 &= \la \Omega,g_i | e^{iH_ft} \left(a_j + a^\dagger_j\right) \left[\prod_{k=0}^{\ell-1} \left(1 - 2a^\dagger_{j+k} a_{j+k} \right)\right] \left(a_{j+\ell} + a^\dagger_{j+\ell}\right) e^{-iH_ft} |\Omega,g_i\ra, \\
                 &= \la \Omega,g_i | e^{iH_ft} \left(a^\dagger_j-a_j\right) \left[\prod_{k=1}^{\ell-1} \left(1 - 2a^\dagger_{j+k} a_{j+k} \right)\right] \left(a_{j+\ell} + a^\dagger_{j+\ell}\right) e^{-iH_ft} |\Omega,g_i\ra.
\end{align}
\end{widetext}
The calculation will proceed more easily if we introduce Majorana fermions, which describe the real and imaginary parts of the Jordan-Wigner fermions,
\begin{equation}
  \gamma^x_\eel = a_\eel + a^\dagger_\eel, \qquad \gamma^y_\eel = -i(a_\eel - a_\eel^\dagger),
\end{equation}
which obey the anticommutation relations $\{ \gamma^a_j, \gamma^b_\eel \} = 2\delta_{a,b}\delta_{j,\eel}$. In terms of these Majorana operators we have
\begin{equation}
  a^\dagger_j a_j = \frac12 \left( 1+ i \gamma_j^x\gamma^y_j\right), \quad \exp(i\pi a^\dagger_j a_j) = -i \gamma^x_j \gamma^y_j,
\end{equation}
and hence
\begin{equation}
  t^{xx}(\ell;t) = \la\la \Omega_e, g_i| e^{iH_ft} \prod_{k=1}^{\ell} (-i\gamma^y_k \gamma^x_{k+1}) e^{-iH_ft} |\Omega_e,g_i\ra\ra.
\end{equation}
This can be written as the Pfaffian of a $2\ell\times2\ell$ matrix
\begin{align}
  t^{xx}(\ell;t) &= (-i)^\ell \text{Pf}(\Gamma) = \text{Pf}(-i\Gamma), \\
    \Gamma &= \left(
    \begin{array}{cccc}
      \Gamma'_{11} & \Gamma'_{12} & \cdots & \Gamma'_{1\ell} \\
      \Gamma'_{21} & \Gamma'_{22} & \cdots & \Gamma'_{2\ell} \\
      \vdots       & \vdots       & \ddots & \vdots          \\
      \Gamma'_{\ell1} & \Gamma'_{\ell2} & \cdots & \Gamma'_{\ell\ell}
    \end{array}
  \right),
\end{align}
where $\Gamma'$ is a $2\times2$ sub-matrix
\begin{align}
  &\Gamma'_{mn} = \left(
    \begin{array}{cc}
      \la\la \gamma^y_{m-n+j}\gamma^y_j\ra\ra_t - \delta_{m,n} & \la\la \gamma^y_{m-n-1+j}\gamma^x_j\ra\ra_t \\
      \la\la \gamma^x_{m-n+1+j}\gamma^y_j\ra\ra_t & \la\la \gamma^x_{m-n+j}\gamma^x_j \ra\ra_t - \delta_{m,n}
    \end{array}
                                                    \right),
\end{align}
where
\begin{equation}
  \la\la O \ra\ra_t = \la\la \Omega_e,g_i| e^{iH_ft} O e^{-iH_ft} | \Omega_e,g_i\ra\ra.
\end{equation}
Using reflection symmetry and translational invariance, we can rewrite this sub-matrix as
\begin{align}
  \bar \Gamma_\eta &= -i\Gamma'_{m+\eta,m}, \\
  &= -i \left(
    \begin{array}{cc}
      \la\la \gamma^y_{\eta+j} \gamma^y_j \ra\ra_t -\delta_{\eta,0} & \la\la \gamma^y_{\eta-1+j} \gamma^x_j\ra\ra_t \\
      -\la\la \gamma^y_{-\eta-1+j}\gamma^x_j\ra\ra_t & - \la\la \gamma^y_{\eta+j} \gamma^y_j \ra\ra_t + \delta_{\eta,0}
    \end{array}
  \right)\\
  &=
  \left(
    \begin{array}{cc}
      -f_\eta & g_\eta \\
      -g_{-\eta} & f_\eta
    \end{array}
  \right),
  \label{barGamma}
\end{align}
where $f_\eta$ and $g_\eta$ can be obtained using Eq.~\eqref{timeEvoJordanWigner}, and they read:
\begin{align}
	\label{fEta}
  f_\eta &= \int_{-\pi}^{\pi} \frac{\rd k}{2\pi} e^{ik\eta} f(k), \\
  f(k) &= i\sin\Big(2\epsilon_{g_f}(k)t\Big)\sin\Big(\Delta_k\Big), \\
	\label{gEta}
  g_\eta &= \int_{-\pi}^{\pi} \frac{\rd k}{2\pi} e^{ik\eta} g(k), \\
  g(k) &= -e^{i\Delta_k} \bigg[ \cos\Big(\Delta_k\Big) - i \sin\Big(\Delta_k\Big) \cos\Big(2\epsilon_{g_f}(k)t\Big)\bigg].
\end{align}
  
When the two-point correlation function is written in terms of~\eqref{barGamma}, it is the Pfaffian of a block Toeplitz matrix:
\begin{equation}
  t^{xx}(\ell;t) = \text{Pf}
  \left(
    \begin{array}{cccc}
      \bar\Gamma_0 & \bar\Gamma_{-1} & \cdots & \bar\Gamma_{1-\ell} \\
      \bar\Gamma_1 & \bar\Gamma_{ 0} & \cdots & \bar\Gamma_{2-\ell} \\
      \vdots       & \vdots         & \ddots & \vdots            \\
      \bar\Gamma_{\ell-1} & \bar\Gamma_{\ell-2} & \cdots & \bar\Gamma_0
    \end{array}
  \right). 
\end{equation}
The matrix here is real and antisymmetric, thus it is an anti-Hermitian matrix. All eigenvalues are imaginary and appear in conjugate pairs. Furthermore, for an antisymmetric matrix the Pfaffian is the square root of the determinant, up to a difficult-to-determine sign. In other words, if an $2\ell\times2\ell$ antisymmetric matrix $\Lambda=-\Lambda^T$ has eigenvalues $\pm \lambda_j$ ($j=1,\ldots,\ell$), the Pfaffian reads $\text{Pf}(\Lambda) = \prod_{j=1}^{\ell} i\lambda_j$.

One can reduce computing the Pfaffian to computing the determinant of an $\ell\times\ell$ matrix $Q$
\begin{align}
	\label{timeEvoTxx}
  t^{xx}(\ell;t) &= (-1)^{\ell(\ell-1)/2} \text{det}(Q), \\
  Q_{nm} &= if_{n-m} + g_{n+m-\ell-1}, 
\end{align}
which is convenient to do numerically. One can also make analytical progress by computing the determinant of the block Toeplitz matrix, as considered in Ref.~[49--51], via applications of Szeg\"o's theorem and Fisher-Hartwig conjectures~[70--76]. 

\subsubsection{The $t^{yy}(\ell;t)$ two-point function}
The computation of this two-point function is very similar to the previous one. In terms of the Majorana fermions,
\begin{equation}
  t^{yy}(\ell;t) = \la\la \Omega_e, g_i| e^{iH_ft} \prod_{k=1}^{\ell} (i\gamma^x_k \gamma^y_{k+1}) e^{-iH_ft} |\Omega_e,g_i\ra\ra.
\end{equation}
The previous expression can be written as the Pfaffian
of a block Toeplitz matrix,
\begin{equation}
  t^{yy}(\ell;t)=
  \text{Pf}
  \left(
    \begin{array}{cccc}
      \tilde\Gamma_0 & \tilde\Gamma_{-1} & \cdots & \tilde\Gamma_{1-\ell} \\
      \tilde\Gamma_1 & \tilde\Gamma_{ 0} & \cdots & \tilde\Gamma_{2-\ell} \\
      \vdots       & \vdots         & \ddots & \vdots            \\
      \tilde\Gamma_{\ell-1} & \tilde\Gamma_{\ell-2} & \cdots & \tilde\Gamma_0
    \end{array}
  \right). \label{tyyToeplitz}
\end{equation}
where $\tilde \Gamma_\eta$ is a $2\times2$ sub-matrix
\begin{equation}
  \tilde \Gamma_\eta  
  =
  \left(
    \begin{array}{cc}
	    -f_\eta & g_{-\eta+2} \\
      -g_{\eta+2} & f_\eta
    \end{array}
  \right),
\end{equation}
where $f_\eta$ and $g_\eta$ are defined in~\eqref{fEta}-\eqref{gEta}.
  
The computation of the Pfaffian is equivalent to the one the determinant of an $\ell\times\ell$ matrix $\tilde Q$
\begin{align}
	\label{timeEvoTyy}
  t^{yy}(\ell;t) &= (-1)^{\ell(\ell-1)/2} \text{det}(\tilde Q), \\
  \tilde Q_{nm} &= if_{n-m} + g_{\ell+3-(n+m)}.
\end{align}

\subsubsection{The $t^{xy}(\ell;t)$ two-point function}
This two point-function can be obtained from $t^{xx}(\ell;t)$. Indeed,  by considering the Heisenberg equation of motion,
\begin{align}
  \frac{\rd}{\rd t} \tau^x_j(t) \tau^x_{j+\ell}(t)
  &=i[H_f,\tau^x_j(t) \tau^x_{j+\ell}(t)]\nonumber\\
  &=2 \bar J \bar g_f \left( \tau_j^y(t) \tau_{j+\ell}^x(t) +\tau_j^x(t)\tau_{j+\ell}^y(t)\right)\,.
\end{align}
By computing the expectation value and considering inversion symmetry we conclude,
\begin{equation}
  t^{xy}(\ell;t)=
  \frac{1}{4 \bar J \bar g_f} 
  \frac{\rd}{\rd t}t^{xx}(\ell;t)\,.
\end{equation}

\section{Details of the numerical algorithm for time-evolution of a driven system}
In this section of the Supplemental Material, we discuss details of the numerical algorithm  used to compute time-evolution of the initial state used in the main text. As we have already shown in the main text, the results of the simulations have an excellent agreement with the analytical results for the integrable case when $\Omega=2 h^z$. We make the following numerical studies outside this integrable line.

We consider the time-evolution of a given initial state $|\Psi_0\rangle$ induced by a time-dependent Hamiltonian, Eq.~\eqref{Ht} of the main text. Formal solution of the time-dependent Schr\"odinger equation gives
\begin{equation}
  |\Psi(t)\rangle = \mathbb{T} \exp\left( -i \int_0^t {\mathrm d}t'\, H(t') \right) |\Psi_0\rangle,
\end{equation}
where $\mathbb{T}$ is the time-ordering operator. Our Hamiltonian $H(t)$ depends smoothly on time $t$, which makes dealing with the time-ordering tricky. To proceed we ``Trotterize'' the time-evolution into $N$ small steps (of size $\Delta t$, such that $N\Delta t = t$), each of which has a fixed Hamiltonian
\begin{align}
  |\Psi(t)\rangle &\approx \mathbb{T} \prod_{j=0}^{N-1} \exp\Big(- i H\big(j\Delta t\big) \Delta t\Big) |\Psi_0\rangle, \\
  &\equiv \mathbb{T} \prod_{j=1}^{N-1} {\cal U}_j |\Psi_0\rangle. 
\end{align}
The problem now becomes how to describe the action of the time-evolution operator ${\cal U}_j$ on a state.

We compute the action of a single step time-evolution operator on a state via the Chebyshev expansion (see, e.g., Ref.~[45]). This avoids the need to diagonalize or exponentiate the Hamiltonian, and it is computationally efficient. The Chebyshev expansion of the time-evolution operator reads:
\begin{equation}
  {\cal U}_j = J_0 \big(\widetilde{\Delta t}\big)\mathbb{1} + 2 \sum_{n=1}^\infty (-i)^n J_n\big(\widetilde{\Delta t}\big) T_n\Big(\widetilde{H}(j\Delta t)\Big). \label{chebyexp}
\end{equation}
Here we have introduced the rescaled time $\widetilde{\Delta t}$ and Hamiltonian $\widetilde{H}(j\Delta t)$, as well as the Bessel functions $J_n(x)$ and the Chebyshev matrices $T_n(x)$, which satisfy the recursion relation
\begin{align}
  T_0(x) &= \mathbb{1}, \quad T_1(x) = x, \\
  T_{n}(x) &= 2x T_{n-1}(x) - T_{n-2}(x).
\end{align}
The rescaled time and Hamiltonian satisfy $\widetilde{H}(j\Delta t)\widetilde{\Delta t} = H(j\Delta t)\Delta t$, and enforce that the eigenvalues of the $\widetilde{H}(j\Delta t)$ lie in the interval $[-1,1]$.

In practice, the Chebyshev expansion~\eqref{chebyexp} has to be truncated to finite order. In the following subsections we present some convergence checks of our ``Trotterization+Chebyshev'' procedure, where we examine the effects of Trotter step size and Chebyshev expansion order. We find that only a low order Chebyshev expansion is required for the problem at hand, leading to a very efficient algorithm for simulating the time-evolution of the continuously (smoothly) driven quantum system. We finish our discussions with a brief inspection of finite system size effects on the nonequilibrium dynamics.

\subsection{Convergence with order of the Chebyshev expansion}
\begin{figure}[t]
  \begin{tabular}{l}
    (a)\\
    \includegraphics[width=0.50\textwidth]{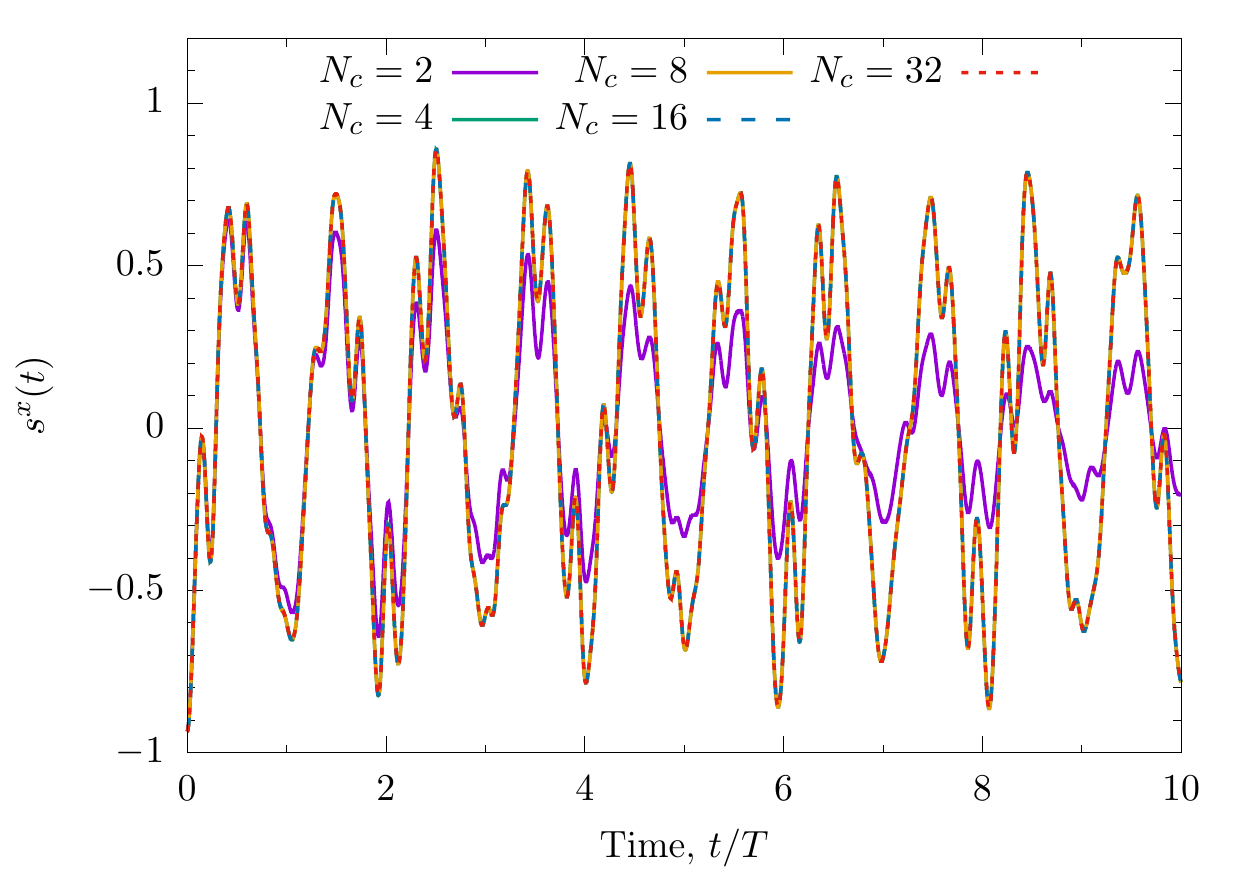}\\
    (b)\\
    \includegraphics[width=0.50\textwidth]{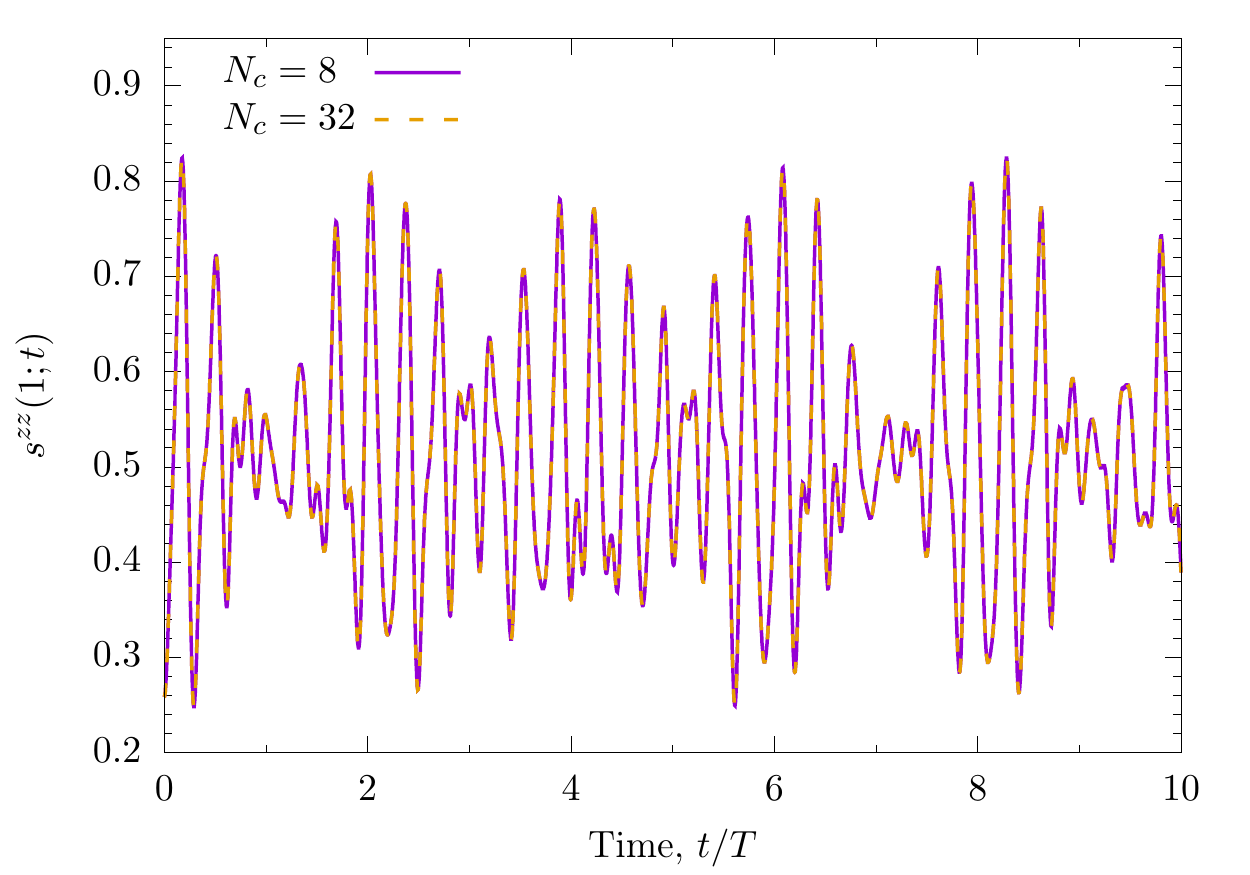}
  \end{tabular}
  \caption{The time-evolution of the one-point function (a) $s^x(t) = \langle \Psi(t) | \sigma^x_1 |\Psi(t)\rangle$; (b) $s^{zz}(1;t) = \langle \Psi(t) | \sigma^z_j  \sigma^z_{j+1} |\Psi(t)\rangle$ starting from the ground state of the Hamiltonian~\eqref{Ht} at $t=0$ with $J=1$, $h^z = 0$, and $h^x=2$ and time-evolved with $J=1$, $h^z=\Omega=1$, and $g=1.5$. The time-evolution is computed via ``Trotterization+Chebyshev'' with Trotter step $\Delta t = 0.005$ and Chebyshev expansion order $N_c$ for ten periods of the drive,  $T=2\pi/\Omega$.  We see that results are rapidly converging with increased  $N_c$.  Note that there is no heating to infinite temperature, despite the fact we are outside the integrable line $\Omega=2 h^z$.} 
 \label{fig:chebyConv}
\end{figure}

Let us first examine how results of the time-evolution vary with the order of the Chebyshev expansion~\eqref{chebyexp}. We fix the Trotter step size $\Delta t = 0.005$  and then proceed to compute an approximation of the time-evolution:
\begin{equation}
  {\cal U}_j \approx J_0 \big(\widetilde{\Delta t}\big)\mathbb{1} + 2 \sum_{n=1}^{N_c} (-i)^n J_n\big(\widetilde{\Delta t}\big) T_n\Big(\widetilde{H}(j\Delta t)\Big), \label{approxChebyExp}
\end{equation}
for given values of $N_c$. We present an example in Fig.~\ref{fig:chebyConv}(a) for the time-evolution of a one-point function following a quench within the disordered (paramagnetic) phase of the quantum Ising chain (details of parameters are given in the figure caption). We observed that there is rapid convergence of the result for increasing order of the Chebyshev expansion: after ten periods of the drive, the results with $N_c = 8$ and $N_c = 32$ agree to seven decimal places. We also stress the fact that, despite being outside the integrable line $\Omega=2 h^z$, there is no heating to infinite temperature.

One may question whether similarly good convergence is observed for more complicated observables. In Fig.~\ref{fig:chebyConv}(b), we present $s^{zz}(1;t)$ for the same quench, where it is apparent that two-point functions also converge similarly well with increasing $N_c$. Within the main body of the text, we have explicitly checked that the obtained results are independent of the expansion order $N_c$. 

\subsection{Convergence with  Trotter step size}
Let us now turn our attention to the size of the time discretization step. 

\subsection{Finite size effects}
With convergence established for Trotter step size and order of the Chebyshev expansion, we finally examine finite size effects in our simulations of the time-evolution. Fixing $\Delta t = 0.005$ and $N_c  =64$, we examine the nonequilibrium time-evolution of $\sigma^x$ following the same quench as in the previous subsections.

\begin{figure}[t]
  \includegraphics[width=0.5\textwidth]{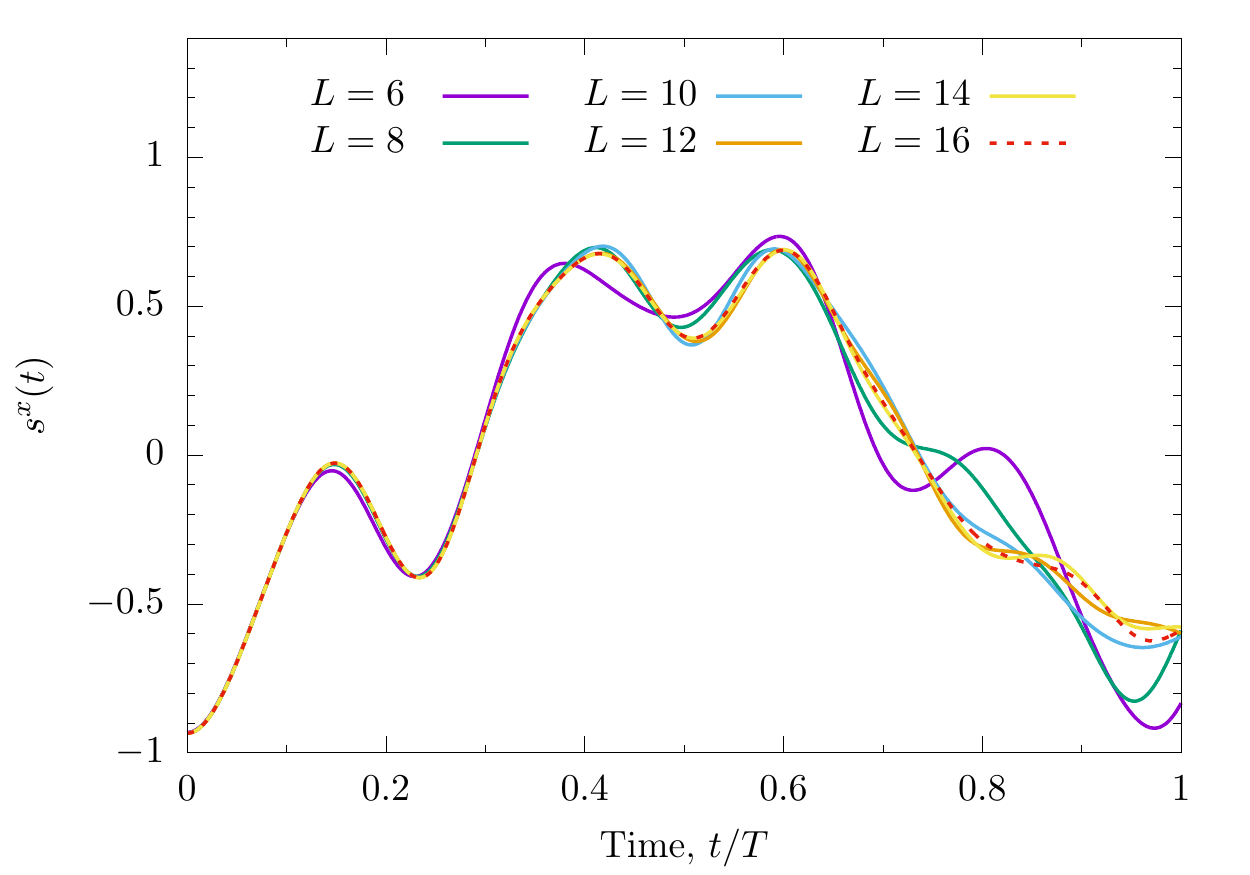}
  \caption{The time-evolution of the one-point function $s^x(t) = \langle \Psi(t) | \sigma^x_1 | \Psi(t)\rangle$ following the quench presented in Fig.~\ref{fig:chebyConv}. Results are presented for a number of system sizes, illustrating finite size effects in numerical data due to small accessible systems. We see that for $L=14,16$ the time-evolution is well matched for almost the whole period of the driving.}
  \label{fig:finitesize}
\end{figure}

We first examine a local observable, $\langle \Psi(t) | \sigma^x |\Psi(t)\rangle$, and check how it behaves with changes in the system size $L$, i.e. what the ``finite size effects'' are. Example data is presented in Fig.~\ref{fig:finitesize}, where we see the expected behaviour: at short times observables for all system sizes are in agreement. Under time-evolution, where propagating excitations are generated, results for different system sizes eventually diverge due to finite size revivals. For a single, sudden quench the picture for this is simple: a quantum quench generates excitations that can propagate around the system. By conservation of momentum, such excitations must be generated in pairs, with momentum $k$ and $-k$. These can propagate around the system, eventually meeting back at the start and interfering -- a purely finite size effect. The time for which this occurs increases linearly with system size. Here we are seeing the driven system analogue of this finite size revival behaviour. Over a single period, we see that results for the two largest system sizes, $L=14$ and $L=16$, match over almost the whole drive period. 

\begin{figure}[t]
  \includegraphics[width=0.5\textwidth]{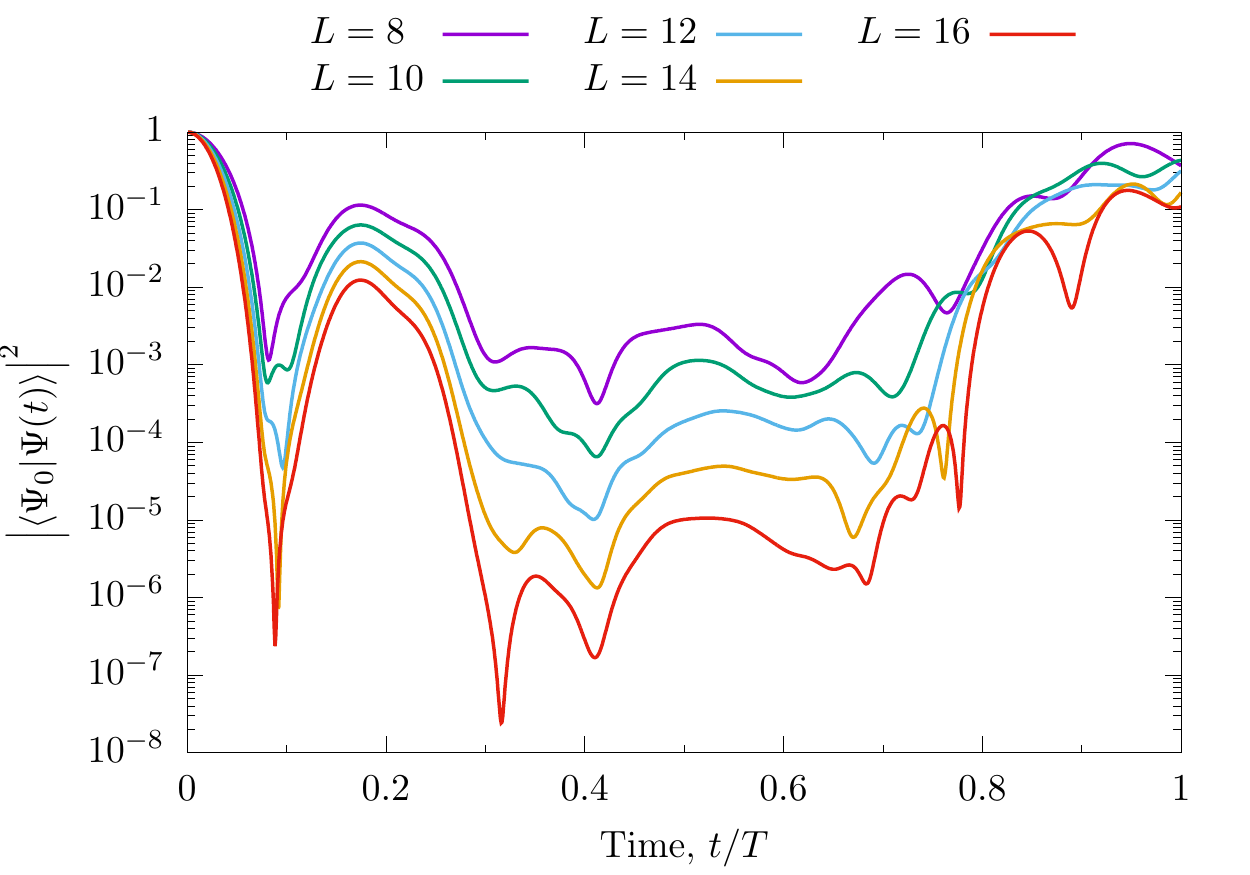}
  \caption{The return amplitude $\Big\vert \langle \Psi_0 | \Psi(t) \rangle\Big\vert^2$ as a function of time $t$ over one period following the same quench as in Fig.~\ref{fig:chebyConv} and Fig.~\ref{fig:finitesize}. Here data is presented for a number of system sizes, $L$, revealing significant finite size effects in this \textit{nonlocal} property. Each simulation was performed with $\Delta t = 0.005$ and $N_c = 64$.}
  \label{fig:finitesizeReturnAmp}
\end{figure}

Finite size effects can more easily be revealed in nonlocal/global measurements of the system. We illustrate this in Fig.~\ref{fig:finitesizeReturnAmp}, where we consider the return amplitude $\Big\vert \langle \Psi_0 | \Psi(t) \rangle \Big\vert^2$ following a quench. This rapidly decays as a function of time towards a close-to-zero value. This close-to-zero value depends \textit{exponentially} on the system size. After one period of driving (the right hand side of the plot) we see that the state $|\Psi(T)\rangle \approx |\Psi_0\rangle$, with the overlap decreasing with system size.

We see that nonlocal observables experiencing severe finite size effects does not carry through to local observables, cf. Figs.~\ref{fig:finitesize} and~\ref{fig:finitesizeReturnAmp}.

\section{Comparison of analytical results with finite volume numerics}
The equations~\eqref{timeEvoTz},\eqref{timeEvoTzz},\eqref{timeEvoTxx}, and~\eqref{timeEvoTyy} for the one and two point-functions are exact in the thermodynamic limit.  However, to compare with our numerical results, it is better to consider analytical expressions for finite size systems. These can be easily obtained by recalling that the integrals in the analytical expressions were obtained by taking the continuos limit of sums over the allowed momenta. Therefore we only have to make the substitution $\int dk \rightarrow (2\pi/L) \sum_{k_n}$ in the corresponding equations. Concretely, we have, the following results for $t^z(t)$ and $t^{zz}(\ell,t)$
\begin{widetext}
\begin{align*}
  t^z(t) &= \frac12 - \frac{1}{L}\sum_n \Big[
  \sin^2\left(\epsilon_{g_f}(k_n)t\right)\sin^2\left(\Delta_{k_n} +
  \frac{\phi_{k_n}}{2}\right) +
  \sin^2\left(\frac{\phi_{k_n}}{2}\right)\cos^2\left(\epsilon_{g_f}(k_n)t\right)\Big],\\
	  t^{zz}(\ell;t) &= - \frac{1}{L^2} \sum_n \sum_{n'}
	  \Bigg\{ e^{i(k_n-k_{n'})\ell} \Big[ \sin(2\epsilon_{g_f}(k_n)t) \sin(\Delta_{k_n})
			  \sin(2\epsilon_{g_f}(k_{n'})t)\sin(\Delta_{k_{n'}})   + \Big( e^{i(k_n-k_{n'})\ell} - 1
  \Big) \nonumber \\ & \qquad \times e^{i(\theta_{k_n}+\theta_{k_{n'}})}
  \Big(\cos(\Delta_{k_n}) - i \sin(\Delta_{k_n}) \cos(2\epsilon_{g_f}(k_n)t)\Big)
  \Big(\cos(\Delta_{k_{n'}}) - i \sin(\Delta_{k_{n'}}) \cos(2\epsilon_{g_f}(k_{n'})t)\Big)
  \Bigg\},
\end{align*}
\end{widetext}
where the set of allowed momenta is,
\begin{equation}
	k_n=\frac{2\pi n}{L}, \quad n=0,\dots,L-1\,,
\end{equation}
while the expressions for $t^{xx}(\ell,t)$ and $t^{yy}(\ell,t)$ can be computed using~\eqref{timeEvoTxx} and~\eqref{timeEvoTyy}, where $f_\eta$ and 
$g_\eta$ are given by the following equations in 
a finite size system,
\begin{align}
  f_\eta &= \frac{1}{L}\sum_n e^{ik_n\eta} f(k_n), \\
  g_\eta &= \frac{1}{L}\sum_n e^{ik_n\eta} g(k_n).
\end{align}

\end{document}